\documentclass[final,1p,times]{elsarticle}
\usepackage{todonotes}
\usepackage[acronym]{glossaries}
\usepackage{array}
\usepackage{cleveref}
\usepackage{tabularx}
\usepackage{csquotes}
\usepackage{booktabs}
\usepackage{multirow}
\usepackage{adjustbox}
\usepackage{url}
\usepackage{enumitem}

\newacronym{refa}{RefA}{Reference Architecture for security-compliant DevOps}
\newacronym{refaap}{RefA-AP}{RefA-based Assessment Protocol}
\newacronym{refaar}{RefA-AR}{RefA-based Assessment Reports}
\newacronym{refaaq}{RefA-AQ}{RefA-based Assessment Questionnaire}
\newacronym{refaaw}{RefA-AW}{RefA-based Assessment Workflow}
\newacronym{refair}{RefA-IR}{RefA-based Improvement Roadmap}
\newacronym{refamr}{RefA-MR}{RefA-based Maturity Report}
\newacronym{refaav}{RefA-AV}{RefA Areas View}
\newacronym{refaartev}{RefA-ArteV}{RefA Artefacts View}

\newcommand*{\rqone}{Does the \acrfull{refa} sufficiently reflect the artefacts ecosystem of DevOps and security compliance?}
\newcommand*{\rqtwo}{Can the \acrfull{refaap} be used to effectively assess security compliance of software development projects following DevOps workflows?}
\newcommand*{\rqthree}{Do \acrfull{refaar} provide useful results?}

\newcommand*{\requirementone}{The solution must enable security compliance in industrial contexts}
\newcommand*{\requirementtwo}{The solution should leverage DevOps principles to enhance risk-based decision-making regarding security}
\newcommand*{\requirementthree}{The solution should clearly demonstrate the security practices employed during the DevOps lifecycle}

\newcommand\greybox[1]{%
  \vskip\baselineskip%
  \par\noindent\colorbox{lightgray}{%
    \begin{minipage}{\textwidth}#1\end{minipage}%
  }%
  \vskip\baselineskip%
}

\newcolumntype{L}[1]{>{\raggedright\let\newline\\\arraybackslash\hspace{0pt}}p{#1}}
\newcolumntype{C}[1]{>{\centering\let\newline\\\arraybackslash\hspace{0pt}}p{#1}}
\newcolumntype{R}[1]{>{\raggedleft\let\newline\\\arraybackslash\hspace{0pt}}p{#1}}

\usepackage{amssymb}
\usepackage{amsmath}

\usepackage{lineno}

\journal{Journal of Systems and Software}

\begin{document}

\begin{frontmatter}

\title{Aligning Security Compliance and DevOps: A Longitudinal Study}

\author[1,2]{Fabiola Moy\'on}
\author[3,4]{Florian Angermeir}

\author[4,3]{Daniel Mendez}

\author[4,3]{Tony Gorschek}

\author[1]{Markus Voggenreiter}

\author[5]{Pierre-Louis Bonvin}

\affiliation[1]{organization={Siemens Foundational Technologies},
            city={Munich},
            country={Germany}}
\affiliation[2]{organization={Technical University of Munich},
            city={Munich},
            country={Germany}}

\affiliation[3]{organization={fortiss}, cite={Munich}, country={Germany}}

\affiliation[4]{organization={Blekinge Institute of Technology}, city={Karlskrona}, country={Sweden}}

\affiliation[5]{organization={University of Applied Sciences}, city={Munich}, country={Germany}}

\begin{abstract}
Companies adopt agile methodologies and DevOps to facilitate efficient development and deployment of software-intensive products. This, in turn, introduces challenges in relation to security standard compliance traditionally following a more linear workflow. This is especially a challenge for the engineering of products and services associated with critical infrastructures. To support companies in their transition towards DevOps, this paper presents an adaptation of DevOps according to security regulations and standards. 

We report on our longitudinal study at Siemens AG, consisting of several individual sub-studies in the inception, validation, and initial adoption of our framework based on RefA as well as the implications for practice. RefA is a prescriptive model of a security compliant DevOps lifecycle based on the IEC 62443-4-1 standard. The overall framework is aimed at professionals, not only security experts, being able to use it on implementing DevOps processes while remaining compliant with security norms. We demonstrate how RefA facilitates the transfer of security compliance knowledge to product development teams. This knowledge transfer supports the agility aim of ensuring that cross-functional teams have all the skills needed to deliver the compliant products. 
\end{abstract}

%% Keywords

\end{frontmatter}

%------------------------------------------------------------
\section{Introduction}\label{sec:intro}
%------------------------------------------------------------
Many companies embrace agile and DevOps principles as a means for efficient development and deployment of software systems \cite{olsson:2014:stairway}. In recent years, an increasing number of IT/OT companies have adopted agile development and, in particularly, DevOps to build products for industrial and automation control systems (IACS) more quickly \cite{johnson:2023:industrialDevOps}, to enable continuous adaptation and improvement of development processes \cite{hasselbring:2019:industrialDevOps}, and to support digitalization. This development. This development highly increases the amount of software such as traditional and new cloud-based products\cite{gorschek:2018:evolution,fassbinder:2022:digitalization}.

Since IT/OT products often operate in sensitive domains, for example critical infrastructure or healthcare, regulators are concerned about cybersecurity attacks. Consequently, regulators require manufacturers to ensure product security throughout the product's engineering lifecycle (see, for example the EU Cybersecurity Resilience Act \cite{EC:2022:CRA}). In this context, companies implement practices dictated by relevant security standards and obtain the certifications to demonstrate compliance \cite{IEC11}. 
This is known as 'normative process improvement', where compliance is achieved by demonstrating that processes and their outcomes (artefacts) fulfill regulatory norms and standards.

However, the adoption of DevOps renders such compliance processes challenging since security standards come with process requirements that, while align well with structured (plan-driven) software process models, are difficult to reconcile with agile development and DevOps methodologies \cite{beck:2001:manifesto, iso:32675:2022, kim:2016:devops}.

In Continuous Software Engineering (CSE), which encompasses continuous improvement influenced by agility and DevOps \cite{Bosch:2014:continuous}, security and compliance are integral to ongoing discussions \cite{Fitzgerald:2017:Continuous}. However, how to adopt DevOps while complying with existing security standards remains a challenge.
Our work on Continuous Security Compliance (CSC)\cite{angermeir:2024:acsc} demonstrates the limited discussion on agile development and DevOps concerning security compliance  in literature~\cite{moyon:2020:mapping, Bitra:2021}. 
Furthermore, our previous work highlights that emerging approaches to security compliance require a focus on security standards applicable to product development and further validation in industrial contexts to demonstrate their feasibility and potential for adoption (see i.e. \cite{akujobi:2021:model,kumar:2020:adoc}).

We have addressed the gap in literature by conducting a series of empirical studies along a regulated industrial context, namely the ecosystem of Siemens Technology. The studies cover several sectors (including medium and small partner companies)~\footnote{Note that in this paper, we refer to the entire context in an abstract manner for reasons of anonymity.}. With 'industrial context', we refer to IT/OT environments in which 'industrial products', those designed for IACS, are developed.

Our aim is to enable product development stakeholders in implementing and assessing compliance with the certifiable and most relevant standard for secure product development in the IACS domain, the IEC 62443-4-1 standard \cite{iec:41:2018}. To that end, we developed a framework that allows practitioners to visualize and track security standard practices in a DevOps-based process. Our framework supports the reproducible and transparent adoption of normative security processes improvement in DevOps.

Note, that the dimension of the herein reported longitudinal research program does not allow for extensive study methods descriptions. Our research program involved 102 practitioners in the development of twenty-six different software-intensive products. Twelve of the products in the industrial context at Siemens AG. 
We therefore report, on studies concisely and highlight practical applications, where possible in a condensed manner.
One hope we associate with our work is to support both the research community concerned with security compliance in DevOps and practitioners alike.

Specifically, we make the following contributions: (1) We describe the background challenges that inspired the creation of the solution framework over five studies. Our framework includes an artefact-centric model, the \textbf{Ref}erence \textbf{A}rchitecture for security-compliant DevOps (‘\acrshort{refa}’), as well as, two utilities: the \acrfull{refaap} and \acrfull{refaar} to enable the practical application of our model in security process assessments. (2) we report on the evaluation of the framework in three case studies, and (3) we reflect on our learnings and key results from applying the framework in a real industrial context.

The rest of this paper is structured as follows. In \Cref{sec:background}, we discuss two studies analyzing the research problem, and summarize related work. \Cref{sec:method} describes the research methodology. In this section, we recapitulate on eight studies conducted to construct and evaluate the framework. We then introduce the framework in \Cref{sec:solution}. In \Cref{sec:evaluation}, we elaborate on the evaluation studies, and in \Cref{sec:results} we report our key findings. We conclude this paper with \Cref{sec:conclusion} by discussing the implications of our research for practitioners and researchers.

%------------------------------------------------------------
\section{Background and Related Work}\label{sec:background}
%------------------------------------------------------------
Our research addresses continuous security compliance (CSC)~\cite{Moyon:2018:towards,angermeir:2024:acsc}. It investigates practices to ensure products and processes adherence to security regulatory requirements following continuous software engineering principles and goals. Since adherence to security norms, i.e., \textit{security compliance} (SC), is a significant aim for product development~\cite{Fitzgerald:2013:scaling}, the present work focuses on SC of product development processes following DevOps settings, in short, DevOps(-based) processes. 

In the following sections, we demonstrate the background analysis that sets the foundations of our work. Later, we discuss related research contributions on the intersection of DevOps, security, and compliance. Other relevant terminology used in this paper is available in the Appendix \Cref{tab:02_relevant_concepts}.

\subsection{Background Problem Analysis at Industrial Context}\label{subsec:problem_analysis}
The research presented in this manuscript is rooted in a problem exploration carried out in our industrial context. In particular, we conducted two dedicated studies as a preparation for the present work (see studies 1 and 2 in \Cref{fig:method}). 

In the first study we explored security challenges associated with agile development and DevOps, including those identified in practitioner reports (e.g. \cite{dora_report} and \cite{gitlab_report}). We identified fifteen challenges categorised into five categories: (i) security in continuous development, (ii) transparency of security value to customers, (iii) efficiency of security implementation, (iv) sharing of security knowledge, and (v) security in CI/CD pipelines. The second study consisted of two roundtable workshops with experts responsible for implementing security in IT/OT products and in agile development and DevOps processes. In this study, we investigated the challenges relevances, priority, root causes, and possible solutions. In both studies , we adopted the Delphi \cite{linstone1975delphi} and Lean Coffee \cite{dalton:2018} methods to (i) ensure anonymity in in-person workshops and (ii) focus on topics deemed relevant for participants. The fifteen prioritized challenges are available in our previous publication~\cite{moyon_2024}. 

In Study 2, the experts envisioned long-term goals (see \Cref{tab:03_challenges}). Participants aimed to adapt traditional processes, that rely on robust security gates, to more continuous process flows like DevOps. This enables development and operations team’s to make decisions regarding secure product development and product security. 

This practitioners' aim set the matter of the present research: (1) how to facilitate the adaptation of secure development processes for DevOps, and (2) how to enable product engineering teams to assess and continuously improve security in their processes. To cover the needs of a \textit{regulated context}, we target a reproducible implementation of DevOps in alignment with regulatory norms and standards demands. Given the \textit{industrial product} development context, such security demands are to be derived from the IEC 62443 IACS's security standards family~\cite{iec:62443:2014}. 

Specifically for our work, we apply the IEC 62443-4-1 standard describing secure product development~\cite{iec:41:2018}. Achieving compliance with the IEC 62443-4-1 standard not only contributes towards certification efforts, but also, to establish secure development practices required by the Cybersecurity Regulation Act~\cite{EC:2022:CRA}.

These two studies prompted a task force to guide approaches for achieving the long-term goals. Our partnership with this taskforce facilitated design and regular validation of our framework to establish and assess SC in DevOps-based processes, or in short, 'security-compliant DevOps'. 
\begin{table}[hptb!]
\centering
\scriptsize
\begin{tabularx}{\textwidth}{X X}
\toprule
\multicolumn{2}{l}{\textbf{Top 3 challenges associated with secure agile development and DevOps}} \\
\midrule
{\emph{Roundtable Workshop 1}} & {\emph{Roundtable Workshop 2}} \\
\midrule
{Perform threat modelling and \mbox{consistently} increment and adapt the modelling throughout sprints} & {Make security architecture visible in the backlog and documentation} \\
& \\
{Prioritize security requirements vs. system functionalities } & {Enable security knowledge and ownership in engineering teams} \\
& \\
{Balance efforts for security: improving process compliance vs. improving product quality} & {Implement security activities into the early feedback principle of DevOps} \\
& \\
\toprule
\multicolumn{2}{l}{\textbf{Root Causes}} \\
\midrule
\multicolumn{2}{p{\textwidth}}{
\begin{itemize}[label=-, nosep, leftmargin=*, topsep=0pt, partopsep=0pt]
\item Limited knowledge of security by business stakeholders or product owners, resulting in decisions that could neglect or miscategorise \newline security matters.
\item Linear development processes that focus on quality gates that block the entire delivery flow if one quality gate closes.
\item Difficulty in mapping security requirements to the product backlog, thereby complicating their prioritization.
\item Limited impact of security awareness initiatives due to the inherent complexity and broad spectrum of security topics.
\end{itemize}} \\
\toprule
\multicolumn{2}{l}{\textbf{Long-term goals to frame solution approaches}} \\
\midrule
\multicolumn{2}{p{\textwidth}}{
\begin{itemize}[label=-, nosep, leftmargin=*, topsep=0pt, partopsep=0pt]
\item Move from quality gates in the development flow to a risk-based approach.
\item Evolve from process thinking to awareness.
\item Pursue automation of security and systematic integration into CI/CD pipelines.
\end{itemize}} \\
\bottomrule
\end{tabularx}
\caption{Summary of Challenges, as presented in \cite{moyon_2024}.}
\label{tab:03_challenges}
\end{table}

\subsection{Related Work}
The research field studying security in DevOps (called 'DevSecOps'~\cite{iso:32675:2022}) is broad~\cite{myrbakken:2017:multivocal}. Contributions span (i) the understanding of practitioners challenges (e.g., \cite{rahman:2016:synthesizing,rajapakse:2022:devsecops:challenges}), (ii) the implementation of traditional security activities (e.g., \cite{jaatun:2018:devops:incident}), or (iii) the automation of security in the CI/CD pipeline or cloud environments (e.g., \cite{kumar:2020:adoc,akujobi:2021:model}). Such contributions address practice challenges, to the extent of even proposing DevSecOps models. However those approaches lack a clear focus on compliance or an analysis in regulated environments. 

For SC for continuous software engineering, literature reviews revealed a limited number of papers ~\cite{moyon:2020:mapping, Bitra:2021,ramaj:2022:holding}. Publications on SC in DevOps investigate (i) how DevOps processes, in regulated environments, can integrate security activities, such as threat modeling, secure coding, or security testing (e.g., \cite{yasar:2016:where,yasar:2017:implementing}), or (ii) considerations for security-compliant DevOps in cyber-physical systems (e.g., \cite{abrahamsson:2020:towards}). Despite the analysis in regulated contexts, these contributions lack attention on compliance with any specific standard. Normative process improvement, targeting compliance with the IEC 62443-4-1 security standard is a gap covered by the herein presented research.

To the best of our knowledge, the research contribution closest to this work is \citet{haverinen:2023:information}, who outline design principles for security-compliant DevOps using an \textit{information-centric approach} based on the IEC 62443-4-1 security standard. While such design principles are valuable, they do not provide support for the engineering teams in conducting security compliance assessments on their own. Empowering engineering teams is in scope of our own ambition. 

More generally, the research directions described above, primarily propose rigid \textit{activity-based} approaches where companies would need to adapt their DevOps processes activities to reflect external best practices and norms (c.f. activity-based vs artefact-based approaches \cite{mendez:2015:artrepi}). During our own research on CSC, we examined such activity-based approaches \cite{daennert:2019:assess:csc} and concluded that, while security practitioners appreciate them, their benefits for engineers are limited. Engineers require SC guidance that fits with concrete outcomes of their DevOps practices and activities, meaning, the process artefacts. An artefact-based guidance serves also a broader group of process stakeholders like business owners, or SC assessment roles like auditors. Discussions with practitioners over the past four years also reaffirmed the utility of abstracting from complex activities and focusing on artefacts to satisfy the needs of engineers.  

Therefore, we adopt an artefact-centric approach that focuses on process outcomes (artefacts) rather than on activities, which may differ from context to context. This approach enhances the transparency of the required SC evidence in context-sensitive DevOps flows. As a fundamental source, we adhere to the IEC 62443-4-1 standard, which artefacts were systematically distilled in our previous work~\cite{Moyon:2018:towards}. Subsequently, we extend the SC use case of external assessors evaluating DevOps security activities with the use case of internal engineers using a blueprint to establish security-compliant DevOps processes.

Our work extends the scope of software security assessment frameworks like OWASP-SAMM \cite{owasp-sammCoreModel} and BSIMM \cite{mcgraw:2015:bsimm}, since our framework systematically adopts a certifiable recognized IACS standard while considering the DevOps artefacts to be used as compliance evidence. Artefacts analysis is outside the scope of these general assessment frameworks. Given the focus on DevOps, \acrshort{refa} systematically integrates recognized grey publications that serves as DevSecOps seminal guidance for practitioners (e.g. \cite{bird2016devopssec,humble2010continuous}). Such publications lack emphasis on security compliance. 
%------------------------------------------------------------
\section{Research Method}\label{sec:method}
\begin{figure}
    \centering
    \includegraphics[width=0.98\linewidth]{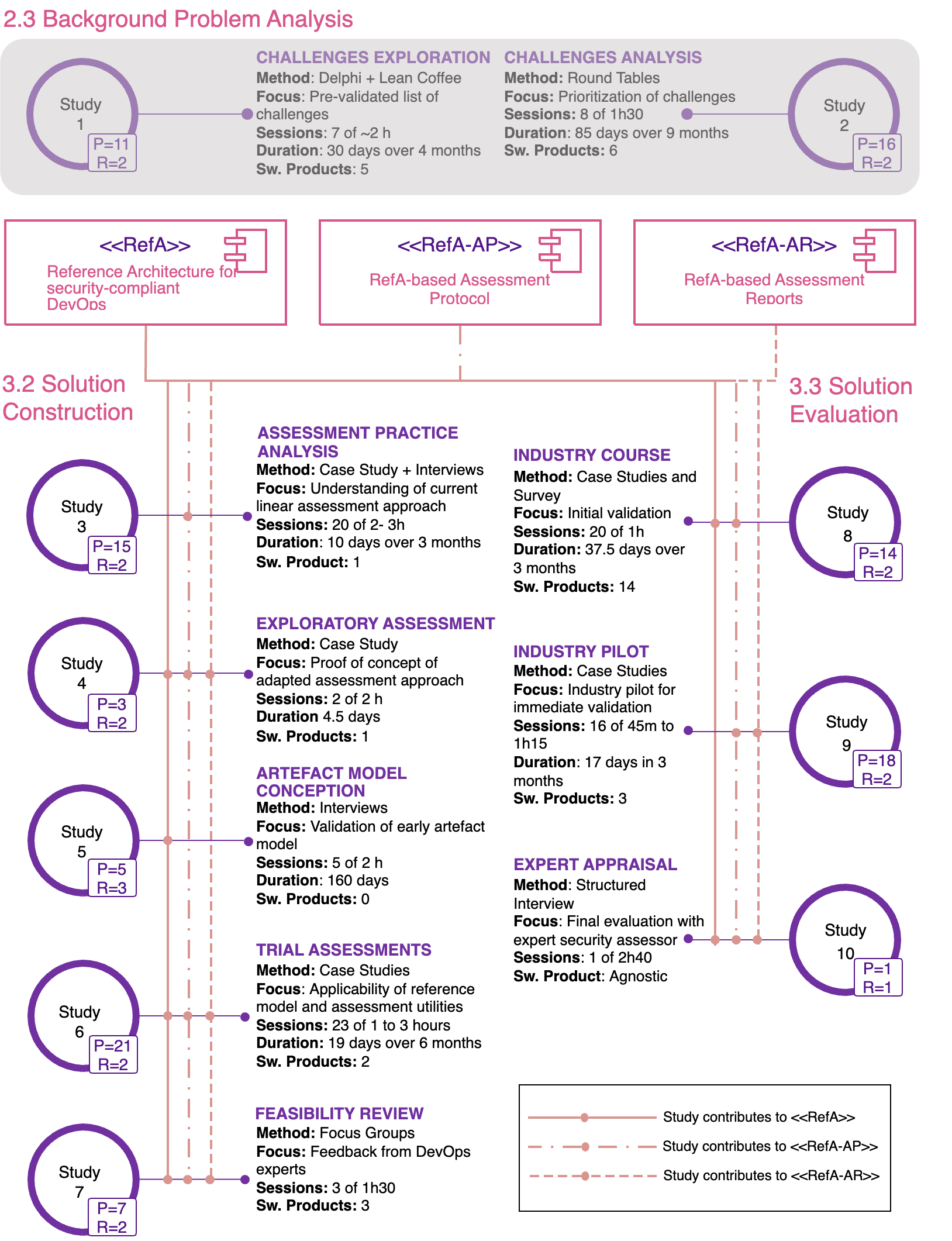}
    \caption{Research Methodology: An overview of the studies conducted to analyse the problem and later construct and evaluate our framework's components for security compliance in DevOps. The rectangles include the number of participants (P) and researchers (R). This figure focuses on how the studies contribute to the components, for a detailed view of the components and their relationship, please refer to \Cref{fig:refa_elements}.}
    \label{fig:method}
\end{figure}

%------------------------------------------------------------
To account for the complexity of the research matter and to ensure validity of our results, we conducted a longitudinal study comprising ten empirical studies over five years (2019-2023). \Cref{fig:method} illustrates the interrelation between the problem analysis and: (i) the design and construction phase of our approach, and (ii) the evaluation phase of the approach. We structured the research program in a co-production fashion~\cite{gorschek2021solving} resembling the principles of action design research~\cite{sein:2022:actionDesignResearch}. We first analysed the problems and challenges in our organisational context to identify the relevant design requirements (see \cref{subsec:problem_analysis}). We then iteratively developed and evaluated the solution, guided by the initially identified industry needs. The construction and evaluation allowed us to gradually scale up from initial validations (proof of concepts, trial runs, and focus groups) to evaluations via industrial pilot studies and expert interviews. Note that the order of the studies themselves within the phases was driven by the opportunities given in our organisational context.

In terms of a philosophical stance, the problem analysis entailed inductive research, where we identified challenges based on experiences and expert opinions. The subsequent phases of the research process entailed a deductive research approach, in which we evaluated the results of our work against the previously identified challenges. Given the qualitative nature of the studies and the strong dependency on the practical experiences of practitioners in the real world, our research methodology exemplifies pragmatic constructivism. \\

In \Cref{subsec:solution_construction}, we describe the incremental design of our framework, and in \Cref{subsec:solution_evaluation} we detail the evaluation. For reasons of complexity and confidentiality, we focus on the key insights gained during the construction studies (Studies 3-7). We present the results of the evaluation studies (Studies 8-10) in \Cref{sec:evaluation}.

\subsection{Solution Design} \label{subsubsec:problemSolutionDesign} 
Motivated by the problem analysis (see \Cref{subsec:problem_analysis}), we designed our solution approach as a framework for product engineering teams, where they can visualize and assess the security status of their agile development and DevOps processes in alignment with security standards. In response to the practitioners’ goals and challenges, we identified the following high level design goals that informed our approach:

\begin{itemize}[label=\textbullet, nosep, leftmargin=*, topsep=4pt, partopsep=0pt]
    \item \textit{Design Goal 1: \requirementthree}. The approach should provide visibility into the implementation of security practices within the DevOps lifecycle, including how these security practices can adopt DevOps principles of automation, shift-left, and continuous assessment.
    \item \textit{Design Goal 2: \requirementone}. The approach is applicable in regulated environments. Consequently, secure product development shall be described according to accepted security standards. Given the specific relevance of the IEC 62443-4-1 security standard for our industrial context and its general relevance in the IACS domain, our approach considers these standard’s practices and artefacts, thereby remaining in compliance with the standard.
    \item \textit{Design Goal 3: \requirementtwo}. The approach should allow for clear indications  of the development process’s security status. Development, operations, security, and business stakeholders should be enabled to discuss risk status and set actions to improvement the DevOps process. This aligns our approach with the DevOps principles of collaboration, incremental development and shift-left.
\end{itemize}

\subsection{Solution Construction}\label{subsec:solution_construction}
Given the design goals, we applied action research to build a framework consisting of three components: a \textit{reference model}, covering design goal 1 and 2, and two \textit{assessment utilities} (Design goal 2 and 3). First, the \acrfull{refa} illustrates in a DevOps lifecycle the security practices extracted from the IEC 62443-4-1 standard. Secondly, the \acrfull{refaap}, and the \acrfull{refaar} use the \textit{model} to allow assessing and reporting compliance with this standard. In \Cref{sec:solution}, we detailed the components of our framework. In this section, we describe in chronological order the series of five studies to incrementally construct the components, see a summary in \Cref{tab:construction_case_studies}.

\begin{table}[htp!]
\centering
\scriptsize
\begin{tabularx}{\textwidth}{X X X p{0.5\textwidth}}
\toprule
\multicolumn{3}{c}{\textbf{Incorporation into the Framework}} & \multicolumn{1}{c}{\textbf{Selected Quotes from Participants}} \\
\cmidrule(r){1-3} 
\textbf{RefA} & \textbf{RefA-AP} & \textbf{RefA-AR} & \\
\midrule
\multicolumn{4}{c}{\textbf{Study 3: Assessment Practice Analysis}} \\
\midrule
Observation of the use of reference models in security compliance assessments & Tailoring of previously contributed security compliance assessment questionnaire \cite{moyon:2020:tool}, adapting it for agile contexts & Eliciting of reporting requirements based on assessors' reporting technique, and the assessees' appreciation of reports & 
\begin{itemize} [label=-, nosep, leftmargin=*, topsep=0pt, partopsep=0pt] 
\item \textit{``It is not possible to gather the team so long for interviews''} Lead Developer.
\item \textit{``What is the name of this security framework?"} Security Expert.
\item \textit{``We want to understand if teams apply security best practices"} Business Owner.
\item \textit{``Questionnaire must remain private to assessors."} Security Consultant.
\end{itemize} \\
\midrule 
\multicolumn{4}{c}{\textbf{Study 4: Exploratory Assessment}}\\
\midrule 
Introduction of a preliminary evaluated activity-based reference model \cite{moyon:2020:devops}. Eliciting practitioners' requirement to observe compliance artefacts precisely & Refactoring of questionnaire structure according to DevOps phases First version of assessment workflow & Design of initial report templates: \textit{compliance heatmap} and \textit{improvement roadmap} &\begin{itemize}[label={--},noitemsep,leftmargin=*,topsep=0pt,partopsep=0pt]
\item \textit{``We should analyse the links from DevSec and SecOps"} Ops Engineer 
    \item \textit{``These activities (code analysis and vulnerability scan of dependencies), we already automated"} Dev Engineer. 
    \item \textit{``Which reports are required by the standard?"} Dev Engineer.
    \item  \textit{``Where are the product and process requirements?"} Security Process Expert.
    \end{itemize}\\
\midrule 
\multicolumn{4}{c}{\textbf{Study 5: Artefact Model Conception}}\\
\midrule 
Introduction of first artefact-based reference model showing security compliance evidence in DevOps \cite{Bitra:2021}. Understanding of benefits and limitations. & Refinement of questionnaire to address artefacts&{N/A}&Extracted from \cite{Bitra:2021}\begin{itemize}[label={--},noitemsep,leftmargin=*,topsep=0pt,partopsep=0pt]
\item\textit{``This is a checklist for engineers to prioritize implementation"} Security Consultant 
    \item \textit{``I would like to map with artefacts in a project, automatically"} Senior Security Standard Expert. 
    \item \textit{``There are far more artefacts than these in DevOps"} Security Consultant.
    \item \textit{``Model needs structural and visual improvement...deeper relationships and consistency."} Security Consultant. 
    \end{itemize} \\
\midrule 
\multicolumn{4}{c}{\textbf{Study 6: Trial Assessments}}\\
\midrule 
Development of \acrshort{refa}'s views after an extension of the artefact-based model and refactoring inspired by the activity-based model. &Structuring of questionnaire according to \acrshort{refa} model. Re-edition of questions and introduction of assessors' guidance. & Use of templates to report assessment results e.g. \textit{compliance heatmap} and \textit{improvement roadmap} &\begin{itemize}[label={--},noitemsep,leftmargin=*,topsep=0pt,partopsep=0pt]
\item \textit{"I see the value of the assessment"} Product Owner 
\item \textit{"I would like to see a hierarchy for artefacts to see which ones must be integrated and which ones not"} DevOps Engineer 
\item \textit{"Prioritization is visible in the report"} 
\textsf{Software Architect} 
\item \textit{``A heatmap, that is a good idea"} Senior Security Processes Consultant
    \end{itemize}\\
\midrule 
\multicolumn{4}{c}{\textbf{Study 7: Feasibility Review}}\\
\midrule 
Introduction of \acrshort{refa} to industry practitioners in engineering and security teams& Design of modular assessment sessions. Establishment of assessment workflow participants and timeframe. &  Introduction of \textit{spider chart} for reporting maturity status of security practices &\begin{itemize}[label={--},noitemsep,leftmargin=*,topsep=0pt,partopsep=0pt]
\item \textit{"Practices is the right word. We should convey to a same maturity model for all continuous practices."} Agility and DevOps Expert 
\item \textit{"I really like it because the testing phase has everything I was expecting"} Agility and DevOps Expert 
\item \textit{"This is very extensive...We cannot make it mandatory for all teams."} Secure Agile Development Expert  
\item \textit{"The questionnaire is agnostic from the framework you use"} Secure Agile Development Expert 
    \end{itemize}\\
    \bottomrule 
\end{tabularx}
\caption{Solution Construction: Overview of studies and their incorporation into the framework}
\label{tab:construction_case_studies}
\end{table}
We analysed the suitability of existing approaches used to assess security in DevOps projects (see Study 3). This analysis led to the development of early versions of \acrshort{refaap} and \acrshort{refaar}. Their subsequent validation during the assessment of a DevOps project revealed that practitioners require a reference model and specific artefacts for DevOps and security implementation (see Study 4). Consequently, we developed and validated an initial artefact model that served to better contextualize the existing version of \acrshort{refaap} in DevOps and agile development settings (Study 5). After calls to disseminate the artefact model, assessment protocol, and reporting method, we assessed a DevOps industry application (Study 6). This pre-validation challenged the completeness of the artefact model, as well as its level of transparency for different stakeholders. Following an improvement phase, we developed the \acrshort{refa} and \acrshort{refaap}. Finally, we interviewed experts to validate the suitability of \acrshort{refa}, \acrshort{refaap}, and \acrshort{refaar} for security compliance assessment of DevOps projects (Study 7). In the following sections, we present the studies and describe the construction of each component.

\paragraph{\textbf{Study 3: Assessment Practice Analysis}}\label{par:study3}
The aim of this study was to examine security assessment methods used to evaluate software development processes, and later adapt them to assess security compliance in continuous software engineering environments. Hence, in 2019, we used an existing assessment method based on the CMM+Secure reference model to assess the security compliance of the development process for a system comprising a web application, a backend, a database system, and a delivery pipeline. 

We conducted interviews with assessors who applied the assessment method both before and after applying it. The organization building the system followed the Scrum framework and included two agile development teams, two operation teams, one security expert, and one service management team. The assessment was conducted by two assessors over three months. The engineering team that participated in the study consisted of two business stakeholders, ten SW engineers, two SW architects, and one security expert. The assessors evaluated two engineering teams during a four-day workshop. The following activities took place: (i) a kick-off session of two hours, (ii) the actual assessment of security in the development process, with eight two-hour sessions, and (iii) a reporting and analysis of findings session of three hours.

In Study 3, we investigated: (i) the challenges faced by practitioners in formulating questions relevant to the effective assessment of security compliance; (ii) the environment and timeline for security process assessments; (iii) practices for reporting; (iv) security activity flows in both linear and agile software development processes.

\paragraph{\textbf{Study 4: Exploratory Assessment}} \label{par:study4}
This study examined the applicability of early versions of the reference model, focusing on \textit{security activities} such as secure design and secure coding, as well as a related questionnaire to assess security compliance. The reference model was improved based on insights gained from Study 3.

The product, selected for analysis, consisted of a cloud-based service and a cyber-physical system. We concentrated on the cloud-service component, which was developed, deployed, and operated by DevOps teams who followed continuous integration, delivery, and deployment practices. In the end of 2019, two researchers interviewed two security experts in the service engineering team who were responsible for the continuous integration, delivery, deployment, and testing. Additionally, a security process expert validated the differences between linear and continuous development processes for this product. We conducted a semi-structured interview with these practitioners, guided by an early version of the reference model (c.f. \cite{moyon:2020:devops}). The model aligned with the DevOps phases and the eight security practices outlined in the IEC 62443-4-1 standard \cite{iec:41:2018}. During each phase, we asked questions selected from the existing assessment protocol (generated after Study 3). This allowed us to determine whether the DevOps process of the cloud-based service had implemented the security process requirements. 

Study 4 demonstrated the usefulness of a reference model in assessing security compliance for development and operations engineers. The study highlighted the limitations of a reference model that illustrates security process requirements or security activities but not their artefacts. For example, secure implementation (SI) in the IEC 62443-4-1 standard resulted in the activity of secure code review, which, in turn, was associated with the artefacts: code standards and code analysis report. During the interview, the engineers emphasized the need for in-depth discussions about the artefacts in their development process and how they are mapped to the IEC 62443-4-1 standard, rather than approaching the activities first and mapping artefacts afterwards.

\paragraph{\textbf{Study 5: Artefact Model Conception}} \label{par:study5}
This study adopted an artefact-based perspective and validated the applicability of a model in assessing compliance with the IEC 62443-4-1 standard.

The model was based on a literature analysis to identify DevOps artefacts in the context of security compliance. We matched the artefacts we identified with those of the IEC 62443-4-1 standard and modelled their interactions in each corresponding DevOps stage. The artefact-based model was later validated in our industrial context, examining its benefits, limitations, and potential usefulness for assessing security compliance with the IEC 62443-4-1 standard. To this aim, three researchers designed and conducted semi-structured interviews with five experienced security practitioners. The modelling procedures and the model’s validation were performed in 2020.

Practitioners reported that this early version of the artefact model was understandable and could be used as a blueprint for assessing security compliance in DevOps. They provided suggestions for quality improvements in the mapping using the IEC 62443-4-1 standard and requested that the model be used for certification purposes. They also requested catalogue descriptions in the model to reduce ambiguity and misunderstanding of certain artefacts. Some practitioners pointed out that a user of the model would require prior knowledge of the standard and DevOps to follow the flow of artefacts in the model.

\paragraph{\textbf{Study 6: Trial Assessments}} \label{par:study6}
Study 6 had the aim of validating the applicability of our improvements to (i) the reference model, consisting of a combination of the activity-based and artefact-based models; (ii) the assessment protocol, which highlighted the required flow of artefacts; and (iii) the improvement roadmap template to report the prioritisation of findings with the impact vs. effort analysis technique~\cite{andersen:2009:root}. In Study 6, we adopted two perspectives: first, security assessment from the development organization’s perspective and second security assessment from a team perspective.

We conducted security assessments of the product development process of two software-intensive products. The products had been developed in DevOps settings by several agile teams. A security consultant and a researcher conducted the two assessments. To assess the first product, the interview sessions followed the sequence of IEC 62443-4-1 practices, as our objective was to report on the product’s compliance with this standard. We interviewed security experts and twelve engineers. To assess the second product, we followed the sequencing of the DevOps stages. The interview sessions included members of an agile team, which consisted of a product owner, a lead architect, four software engineers, and a security expert. 

In both assessments, we applied excerpts from the assessment protocol and utilized our existing reference models to question the interviewees. We applied the pre-defined reporting structures to report our results. Note that, at the time, the utilities of our framework were in the trial stage. However, the improvement roadmap report was adopted to prioritize the findings of the assessments. The assessments were conducted in 2021. We organized workshops with the engineering teams of 40 and 32 hours each, respectively, with the following components: (i) a kick-off session, (ii) the evaluation of security activities, (iii) a prioritisation of findings session, and (iv) a closing session to review assessment findings with management.

The results of the study demonstrated the efficacy of using an assessment questionnaire and an improvement roadmap for the security compliance assessment of industrial products. We observed that a combination of activity-based and artefact-based models proved useful for software engineers by reducing ambiguity in security compliance aspects and enabling discussions of artefacts as evidence of compliance.

\paragraph{\textbf{Study 7: Feasibility Review}} \label{par:study7}
This study introduced the latest versions of the reference model, \acrshort{refa}, and the associated assessment utilities to practitioners.
In a focus group with senior practitioners, we gathered feedback about applying our solution in real DevOps projects. In 2022, we gathered seven practitioners in three group sessions. They reviewed and discussed: (i) the reference model, (ii) the assessment questionnaire and its relationship to the model, and (iii) the reporting protocol based on the improvement roadmap report. The deliverables were well received by the participants, leading to version stabilization. Moreover, the study allowed the inclusion of a maturity status template in the reporting protocol.

\subsection{Solution Evaluation} \label{subsec:solution_evaluation}
In this section, we describe the evaluation studies for the \textit{reference model} \acrshort{refa} and the \acrshort{refa}-based \textit{assessment utilities} \acrshort{refaap} and \acrshort{refaar}. The evaluation results are presented in \Cref{sec:evaluation}. The evaluation was guided by three research questions based on the specifications determined in \Cref{subsubsec:problemSolutionDesign}; each of which addressed a different part of the proposed solution:
\begin{itemize}
    \item [\textbf{RQ1}] \textbf{\rqone} \\ We evaluated \acrshort{refa} to determine whether it adequately captures security practices in the context of DevOps used in software development projects that employ agile and DevOps principles, thereby satisfying \textit{Design Goal 1}.
    \item [\textbf{RQ2}] \textbf{\rqtwo} \\ With this RQ, we analysed the applicability of \acrshort{refa} in the context of security process assessments. We evaluated \acrshort{refaap} as a key tool for collecting feedback on the compliance status of security practices in DevOps processes. We focused on specific features of \acrshort{refaap}, including its simplicity, understandability, goal orientation, and overall suitability for use in the security assessment of industrial products. This research question responds to \textit{Design Goal 2}.
    \item [\textbf{RQ3}] \textbf{\rqthree} \\ To assess whether the solution adequately deals with the challenges associated with identifying security risks in continuous product development (fulfilling \textit{Design Goal 3}), we evaluated the effectiveness of a \acrshort{refa}-based assessment. Specifically, we examined whether it enabled software engineering teams to clearly visualize missing or incomplete security practices in their development processes while they simultaneously designed improvement roadmaps to elevate security from the current state of the product to a defined target status.\\
\end{itemize}
To answer the research questions listed above, we conducted studies in different contexts, including an industry course (Study 8), an industry pilot project (Study 9), and an expert appraisal (Study 10).

\paragraph{\textbf{Study 8: Industry Course}}
For the initial validation of \acrshort{refa} and its utilities, we partnered with a professional master's degree program offered at Blekinge Institute of Technology and embedded Study 8 in a secure DevOps course given to industry practitioners. We selected this context to ensure that the course participants were practitioners actively working on DevOps projects, independent of our context at Siemens. 

The course participants were introduced to the foundations of agile development and DevOps, as well as the IEC 62443-4-1 standard. After this introduction, the participants were trained to use \acrshort{refa} and \acrshort{refaap} in 20 one-hour sessions in 2022. 
The practitioners were given assignments on the usage of \acrshort{refa} allowing us to gather feedback on the proposed framework. We discuss two of these assignments.

The first assignment was designed to evaluate the applicability of the reference model \acrshort{refa} in guiding security assessments of existing pipelines. To this end, we established an example DevOps project, with participants assuming two roles: first a DevOps engineer who had to integrate security activities into the DevOps project, and second, of a security assessor using \acrshort{refa} to identify deltas of the implementation with respect to \acrshort{refa}’s areas and artefacts. Participants were guided through the \acrshort{refa} to enable an estimation of how well it was understood and applied. Afterward, we collected feedback on benefits and drawbacks and the usefulness for real world usage. The relevant parts of the assignment can be found in \cite{this_online_material}. 

The second assignment focused on the real world application of \acrshort{refa} and \acrshort{refaap}. The participants were asked to assess their own industrial DevOps development projects with \acrshort{refa} and \acrshort{refaap}, and submit their assessment with anonymized data, ensuring that sensitive information was not disclosed. This enabled us to review how well \acrshort{refa} and \acrshort{refaap} were understood and to what extent the practitioners had applied them in their DevOps projects.
The survey contained demographic questions about the practitioners and the context in which they had applied \acrshort{refaap}, as well as open-ended and closed-ended questions to gather feedback on \acrshort{refa} and \acrshort{refaap}. A complete overview of the practitioners’ survey responses can be found in \cite{this_online_material}. Participants were also asked to voluntarily state their general perception of \acrshort{refa}. For unbiased participation, responses were recorded anonymously and dropouts were not considered to avoid an implicit incentive to participate in the survey.

\paragraph{\textbf{Study 9: Industry Pilot}}
We performed an evaluation of \acrshort{refa}, \acrshort{refaap} and \acrshort{refaar} in three product development projects. As part of a broader assessment of CSE practices initiated by product development stakeholders, we collaborated with several agile and DevOps transformation experts within the company to evaluate \acrshort{refa} and \acrshort{refaap} with respect to their security compliance. The assessment plan is available online \cite{this_online_material}.

The assessment of security compliance involved an assessor, the product development team (including individuals responsible for security at the management and technical level), and an external person representing the assessment stakeholders. The assessment occurred in online meetings, allowing the product development team to visualize the \acrshort{refaap} questions and the assessor’s guidance. A shared screen depicted digital copies of \acrshort{refa} so particpants could refer to it for clarifications on the assessment questionnaire.
The team members answered an excerpt of \acrshort{refaaq} in the round. They discussed examples, which they rated according to their knowledge and role responsibility using the \acrshort{refamr} maturity scale. This approach allows us to compare the answers of the security professionals with those professionals who did not have a security background. The assessor’s responsibilities were to clarify questions asked by the assessees and request further evidence, if necessary. The external person oversaw the assessment process to ensure that the questionnaire protocol was being followed, thereby maintaining a high quality of assessment.

Once the assessment was concluded, we presented the outcomes using \acrshort{refa}-based assessment reports within the context of the overarching continuous software engineering assessment process to business owners, product development teams, the responsible security team, and other stakeholders. The results of the assessment were used by business owners and the responsible security team in their deliberations on whether to use \acrshort{refa}, and by the project teams in planning future security improvement work.

\paragraph{\textbf{Study 10: Expert Appraisal}}
This study served as a final appraisal before the approach was released. To obtain a critical comparison between established initiatives and ours, we conducted a guided interview with a security compliance assessor who partake in the development and maintenance of our partner company's security frameworks. The expert addressed the accuracy of the reference model, the completeness of security compliance elements of the IEC 62443-4-1 standard, and determined the applicability of the assessment utilities by highlighting their benefits and shortcomings. The expert was a responsible for product security and possesses knowledge in product quality assessment and development. 
The expert regularly conducts security compliance assessments. The security expert works as an ambassador for the security framework in our industrial context. During the interview, we first introduced \acrshort{refa}, and later \acrshort{refaap} and \acrshort{refaar}. The arrangement of the interview allowed the expert to examine our constructs for up to 30 minutes each. To compare the feedback with that received from other sources, we applied the same survey protocol used by DevOps engineering practitioners during their industry course (Study 8). This allowed us to directly compare the expert’s ratings with those of the course participants across several metrics.
%------------------------------------------------------------
\section{Approach for security-compliant DevOps} \label{sec:solution}
\begin{figure}[htbp!]
 	 	\centering
 	 	\includegraphics[width=\textwidth]{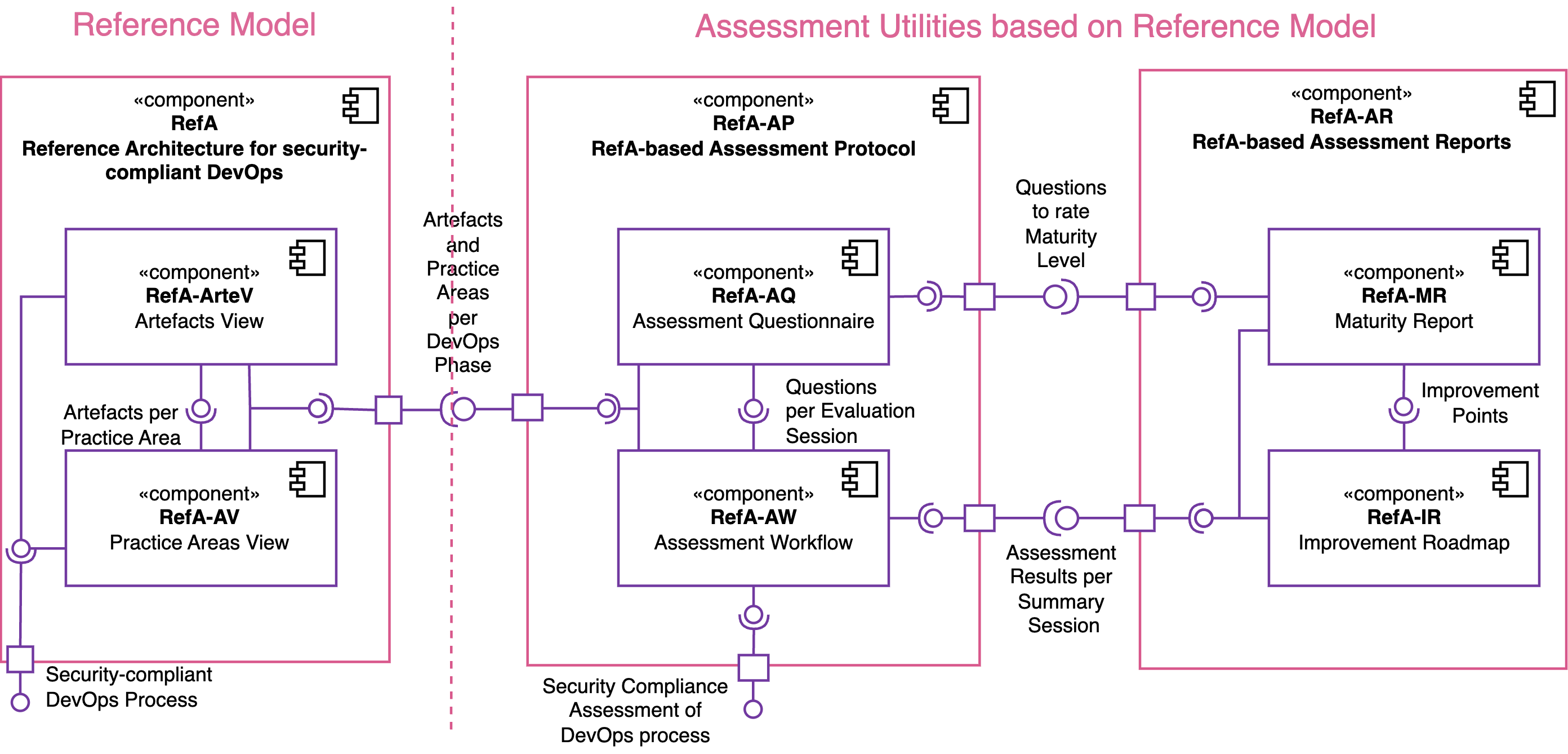}
 	 	\caption{Framework Overview: UML Component Diagram showing the elements of our approach for visualisation and assessment of security compliance in DevOps. The reference model \acrshort{refa} is presented on the left of the figure, and the assessment utilities \acrshort{refaap} and \acrshort{refaar} are presented on the right.} 
 	 	\label{fig:refa_elements}
\end{figure}

%------------------------------------------------------------
Our framework enables engineering teams to adapt their DevOps process to comply with the IEC 62443-4-1 standard. Using \textit{\acrshort{refa}}, the \textit{model}, practitioners visualise a security-compliant DevOps process. By applying the \textit{utilities}: \textit{\acrshort{refaap}} and \textit{\acrshort{refaar}}, professionals can identify existing and missing security practices in their DevOps processes, assess their maturity, and propose improvement actions. In this section, we describe our framework depicted in \Cref{fig:refa_elements}.

\subsection{RefA - Reference Architecture for security-compliant DevOps} \label{subsec:refa}
The \textbf{Ref}erence \textbf{A}rchitecture for security-compliant DevOps: \textbf{\acrshort{refa}} describes a DevOps life-cycle that is compliant with the IEC-62443-4-1 standard~\cite{iec:41:2018}. \acrshort{refa} consists of \textit{view models} depicting security practices extracted from these standard mapped with agile/DevOps practices, i.e. continuous integration and continuous delivery. \acrshort{refa} also includes security practices identified through literature review and practice observation. This helps to overcome limitations of the IEC 62443-4-1 standard to cover Ops. 

\begin{figure}
    \centering
    \includegraphics[width=0.85\textwidth]{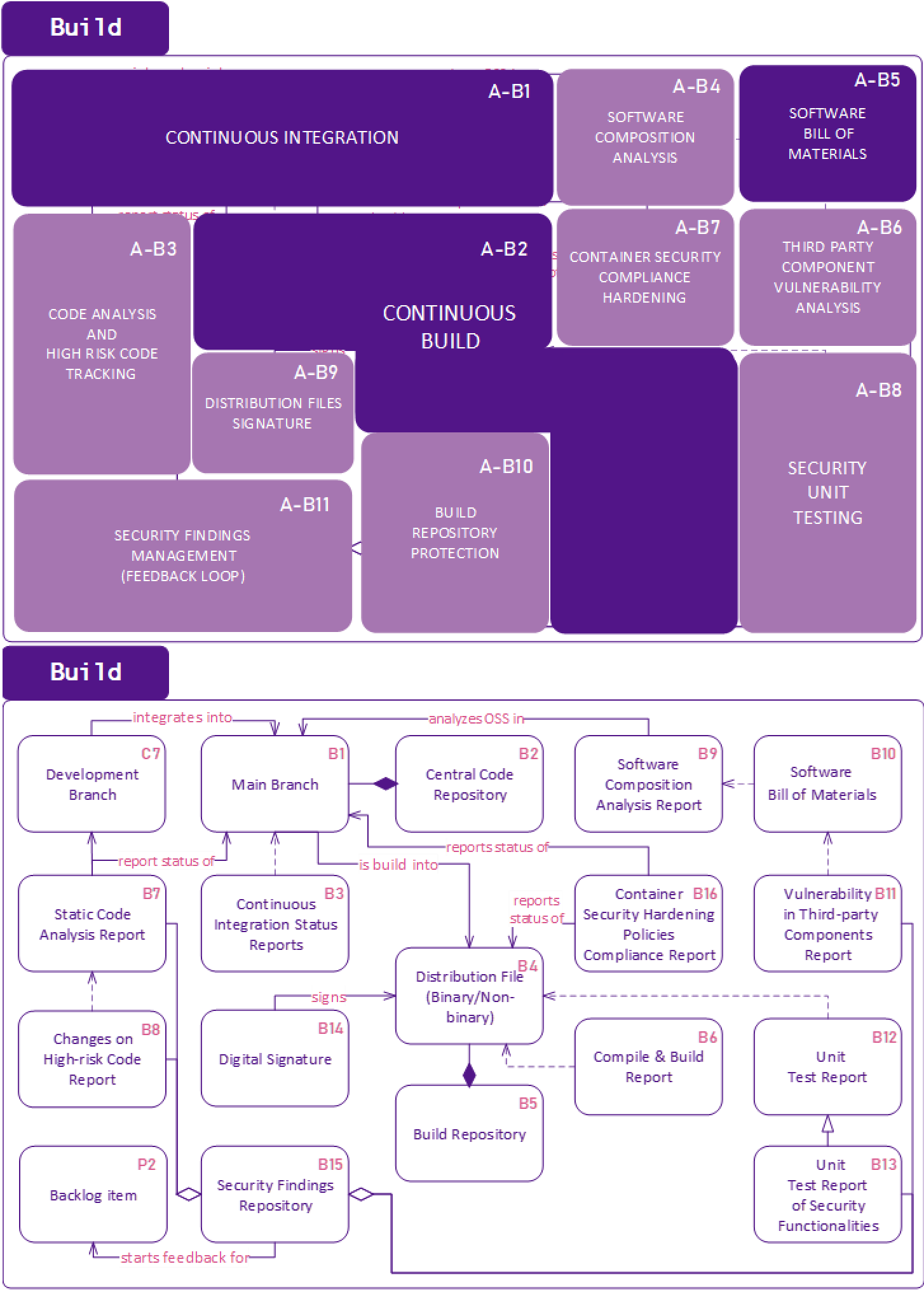}
    \caption{The DevOps Phase Build according to RefA: The model at the top presents software engineering and secure software engineering practices in dark and light purple respectively. The model at the bottom shows the artefacts that serve as evidence of security compliance for the individual practices. A complementary narrative is available in the RefA Technical Report \cite{moyon:2023:refa}.}
    \label{fig:dso_build}
\end{figure}

\acrshort{refa} utilizes the widely adopted DevOps lifecycle, which comprises the following eight phases: \textit{Plan, Code, Build, Test, Release, Deploy, Operate,} and \textit{Monitor} \cite{Pennington:2019}. This approach enables product engineers to easily relate to their current development lifecycle. Within each DevOps phase, \acrshort{refa} offers two representations of security compliance in DevOps: the \textit{artefacts view} \acrshort{refaartev}, which shows the inputs and outputs from DevOps flows in UML notation, and the \textit{practice areas view} \acrshort{refaav}, which consolidates the artefacts into practices, highlighting the overarching software engineering and secure software engineering domains, in dark and light purple respectively. \Cref{fig:dso_build} portrays the DevOps phase \textit{Build}~\cite{moyon:2023:refa}. 

The artefact-centric view of \acrshort{refa} increases the security awareness of agile/DevOps practitioners by highlighting the mapping of IEC 62443-4-1 security standard artefacts and the DevOps flow, as well as DevOps-native artefacts. The area-centric view, on the other hand, reduces complexity by illustrating security practices, the DevOps phase in which they should be implemented, and related agile and DevOps practices. This is required by stakeholders, including business owners, quality engineers, and security process consultants.

\acrshort{refa} is fully detailed in a technical report (c.f. \cite{moyon:2023:refa}), in this paper we provide a summary of the lifecycle in the appendix.

\subsection{RefA-based Assessment Protocol \acrshort{refaap}} \label{subsec:refaAssesProtocol}
This \textit{utility} guides the execution of security compliance assessments of DevOps-based development processes. \textit{\acrshort{refaap}} consists of two subcomponents: \acrshort{refa}\textit{-based Assessment Questionnaire} (\textit{\acrshort{refaaq}}) to validate DevOps processes adherence with practices described by \acrshort{refa}, and \acrshort{refa}\textit{-based Assessment Workflow} (\textit{\acrshort{refaaw}}) to organize efficient assessment workshops (see \Cref{fig:refa_elements}).

\subsubsection{The Assessment Questionnaire \acrshort{refaaq}}
Implemented as a spreadsheet, \textit{\acrshort{refaaq}} contains fifty-eight questions grouped according to \acrshort{refa}'s DevOps Phases and practice areas. Assessors and assessees can map \acrshort{refaaq} to the corresponding \acrshort{refa} practice area and identify which artefacts are required as compliance evidence. Questions found here target the implementation and effectiveness of continuous security practices. For \textit{reproducibility} of the security compliance assessments, each question is motivated and further explained in the \textit{assessor's guide}. Finally, to provide \textit{transparency}, assessors can document the current security status and improvement points in the corresponding columns. \Cref{tab:assessmentQuestionnaire} in the appendix, shows an excerpt of the \textit{assessment questionnaire} in \acrshort{refaap}, focusing on the Code, Build, and Test phases.

\subsubsection{The Assessment Workflow \acrshort{refaaw}}
\acrshort{refaaw} defines the most effective way to conduct security compliance assessments. A assessment can be completed in one sprint and consists of six sessions: four dedicated to the \textit{evaluation}, and two serving as \textit{summary} sessions. The \textit{evaluation} sessions should be held in the first two weeks of the sprint and last 2 hours each. Each evaluation session covers \acrshort{refaaq} questions related to two DevOps Phases: (1) Plan and Code, (2) Build and Test, (3) Release and Deploy, and (4) Operate and Monitor. During these sessions, assessors use \acrshort{refaaq} to determine the compliance status of the process while analysing the security practices described in \acrshort{refa}. Assessors ask for compliance evidence that relates to the artefacts described in \acrshort{refa}'s artefacts view (see \Cref{fig:dso_build}). Assessors note non-compliant practices and conclude each session by listing \textit{improvement points} for each area. The \textit{summary} sessions last 90 minutes and can be conducted during the last weeks of the sprint. The objectives of the sessions include prioritizing the identified \textit{improvement points} and discussing security \textit{maturity} in DevOps. The outcome of the sessions is summarized in \acrshort{refaar} \textit{assessment reports}. 

\subsection{RefA-based Assessment Reports \acrshort{refaar}}\label{subsec:refaAssesReports}
This \textit{utility} covers assessment reporting. \textit{\acrshort{refaar}} visualizes the results of a security assessment across two reports: \acrshort{refa}\textit{-based Improvement Roadmap} (\textit{\acrshort{refair}}) showing the action plan, and \acrshort{refa}\textit{-based Maturity Report} (\textit{\acrshort{refamr}}) depicting the state of implementation of security compliant practices in a DevOps phase  (see \Cref{fig:refa_elements}). The use of \acrshort{refaar} fosters stakeholders collaboration to define plausible action plans and hold transparency regarding the current security status of their development processes.

\subsubsection{Improvement Roadmap \acrshort{refair}}
This report template is an instance of the \textit{impact vs. effort matrix} conceived by \citet{andersen:2009:root}, and adopted as a quality improvement tool \cite{sixsigma:2022:howTo, asq:quality}. The \textit{orthogonal matrix} includes four quadrants: (i) \textit{Quick Wins}, the upper left area where low effort produces a significant impact; (ii) \textit{Major Projects}, the upper right area where significant impact requires major endeavours; (iii) \textit{Fill-ins}, the lower left area where low effort produces insignificant impact; and (iv) \textit{Thankless Tasks}, where high effort delivers insignificant impact. 
\acrshort{refair} is used during the \textit{summary sessions} described in \acrshort{refaaw} (see \Cref{subsec:refaAssesProtocol}). Assessees distribute the \textit{improvement points} in the respective quadrants. The matrix provides a clear visualization of how to prioritize the solutions to security practices that are not in compliance. Stakeholders are encouraged to first pursue the \textit{Quick Wins}, if possible in the next sprint. Thereafter, they should plan how to achieve the \textit{Major Projects}, aided by the Scrum Master, Product Owner, and other Business Stakeholders. Finally, the matrix motivates teams to limit their attention to \textit{Fill-ins}, and, discuss the effect of not working on the \textit{Thankless Tasks}.  

\subsubsection{Maturity Report \acrshort{refamr}}
This report template is an instance of the \textit{radar chart} conceived by Kiviat \cite{Kolence:1973:software}, which depicts \textit{variables} with \textit{radii} and allows for the comparison of several \textit{curves} in the same chart. \acrshort{refamr} displays two curves: the \textit{current status} showing the compliance snapshot to date, and the \textit{target status} showing the compliance status that the process stakeholders envision as their goal. This serves the purposes of plausible planning and expectative management. In \acrshort{refamr}, the \textit{radii} represent maturity levels, according to the ad-hoc maturity scale presented in \Cref{tab:maturityScale} in the Appendix. The maturity scale combines existent maturity approaches, namely: the agile assessment model by \citet{Tuncel:2021:scope}, the continuous assurance flow by \citet{fassbinder:2021:continuous}, the crawl-walk-run methodology applied for agility maturity by \citet{netmind:2021:crawl}, and our observations from practice. The criteria that determine maturity levels include the existence of security practices, the quality of implementation, and the adherence to DevOps principles. Finally, in \acrshort{refamr}, the variables are the security practices depicted in \acrshort{refa}. This enables two levels of granularity while reporting, first Security Maturity per DevOps Phase, and second Security Maturity per DevOps lifecycle. 
%------------------------------------------------------------
\section{Evaluation} \label{sec:evaluation}
%------------------------------------------------------------
We evaluate our framework in three phases. The first phase comprised an initial validation of the \acrshort{refa} and \acrshort{refaap} with industry practitioners (Study 8). The second phase consisted of an evaluation of \acrshort{refaap}, and \acrshort{refamr} within a pilot assessment study (Study 9). The third phase was the final pre-release evaluation of \acrshort{refa}, \acrshort{refaap}, and \acrshort{refamr} by an expert in the field of secure development of industrial products (Study 10).

\subsection{Validation in an Industry Course (Study 8)}
This study focused on evaluating \acrshort{refa} and an assessment using \acrshort{refaap} and \acrshort{refamr}. Study 8 and its two parts were introduced in \Cref{subsec:solution_evaluation}.
\subsubsection{Part 1: Guided Assessment}
\paragraph{Goal}
Study 8 aimed to examine practitioners’ perceptions regarding their application of \acrshort{refa} in DevOps projects. The practitioners were asked to: (i) identify the benefits and drawbacks associated with \acrshort{refa}, (ii) report on their experience of applying \acrshort{refa}, and (iii) determine whether they would apply \acrshort{refa} in real industry projects.

\paragraph{Results}
Out of the fourteen practitioners, eleven gave their perception on \acrshort{refa}. In the following, we elaborate their answers, starting with the observed benefits and drawbacks of \acrshort{refa}, followed by the practitioners’ experience of applying \acrshort{refa}, and concluding with their perception of \acrshort{refa}’s usefulness for the industry.

\paragraph{Benefits of \acrshort{refa}} The benefits of the \acrshort{refa} can be structured into three categories:
\begin{itemize}[label=\textbullet, nosep, leftmargin=*, topsep=4pt, partopsep=0pt]
   \item Structuredness: \acrshort{refa}'s models and description are well-suited to be understood by engineers and other stakeholders. The course participants perceived the structuredness of \acrshort{refa} as positive, it provides a comprehensive overview of activities and stages, clearly distinguishes between security and non-security activities, and offers a reference for an ideal security state in a product. This is achieved by providing a \enquote{thorough mapping of security activities to DevOps phases}, thereby enabling a systematic approach to addressing security in DevOps workflows.
   \item Flexibility: \acrshort{refa} enables flexibility in identifying security aspects within DevOps. \acrshort{refa} facilitates identifying key security activities in DevOps phases to improve the security state through actionable guidance. \acrshort{refa} also enables and facilitates shifting-left security activities as such activities remain visible across the entire DevOps workflow.
   \item Effectiveness/ease of use: \acrshort{refa} is easy to use and helps in understanging security practices in DevOps. \acrshort{refa} is well understandable since it provides a comprehensive list of measures that can be used to improve security in the DevOps workflow. Employing \acrshort{refa} is uncomplicated as it can be read as a 'step-by-step reference to an ideal state'.
\end{itemize}

\paragraph{Drawbacks of \acrshort{refa}} Course participants reported two major issues with \acrshort{refa}. First, they perceived \acrshort{refa} as complex and overwhelming, especially in the beginning. This complexity can make it challenging to follow through with the entire \acrshort{refa}. Second, the participants suggested that additional knowledge might be helpful for non-security practitioners, as \acrshort{refa} contains a large number of acronyms and technical terms. Additionally, one participant argued that domain knowledge specific to each of the DevOps phases is necessary. Another participant mentioned the need for extensive documentation on each area and activity in the different DevOps phases.

In addition to these observations, the participants mentioned issues unrelated to \acrshort{refa}. For example, users of \acrshort{refa} could interpret activities and artefacts differently, \acrshort{refa} cannot cover every possible scenario, and that practitioners might misinterpret \acrshort{refa} as instantiating a \enquote{gold standard} and therefore limit themselves to only what is prescribed in \acrshort{refa}, rather than addressing the security needs in their specific use-case.

\paragraph{Experience in applying \acrshort{refa}}
Overall, the participants’ experience of using \acrshort{refa} was positive. They highlighted \acrshort{refa}’s ease of use in security coverage and in identifying missing security activities. The level of detail offered by \acrshort{refa} enables it to serve as a reference book, highlighting the steps towards achieving an improved security state. For one participant, \acrshort{refa} helped conceptualise an ideal DevOps pipeline. Another participant mentioned \acrshort{refa} only seems challenging in beginning, but when correctly applied, \acrshort{refa} is very efficient and facilitates the work done by DevOps teams. One user mentioned that inexperienced users might become confused when mapping security activities with tools, as these tools can cover more than one security area. The participant who found \acrshort{refa} helpful in envisioning an ideal DevOps pipeline mentioned that they still lacked knowledge about minimal pipeline setups and experienced difficulties in prioritizing security activities based on \acrshort{refa}.

\paragraph{Usability of \acrshort{refa} for industry}
When asked if they would use \acrshort{refa} to assess the extent to which a DevOps pipeline implemented the security areas of RefA, eight out of eleven practitioners said that they would. They deem \acrshort{refa} to be a powerful tool to evaluate security coverage quickly. They valued its comprehensiveness and coverage of the most important security areas in practice. They also see \acrshort{refa} as a meaningful way to \enquote{bridge the gap between theory and real implementation}, thereby facilitating communication between stakeholders. Two of the eleven participants stated that they will use parts of \acrshort{refa}, prioritizing and cherry-picking the security practice areas they deem necessary and valuable. For both participants, the decision was primarily based on limited project resources that hindered a complete adoption of the security practice areas described in \acrshort{refa}. Even though they were motivated to use \acrshort{refa}, one participant claimed that mapping activities to their distributed pipeline setup would take some effort.

\subsubsection{Part 2: Survey-based Self-assessment}

\paragraph{Goal}
The second part of Study 8, aimed to understand practitioners perceptions of \acrshort{refaap} across thirteen metrics (see \Cref{tab:eval_study_2_assessment}). To prevent the reporting of neutral ratings, the participants’ responses followed a Likert scale from 1 (‘I strongly disagree’) to 8 (‘I strongly agree’). We also solicited practitioners opinions about \acrshort{refa} across five metrics (see \Cref{tab:eval_study_2_refA}). The course participants applied \acrshort{refa} and \acrshort{refaap} to a real project they were working on. This part of the survey was not anonymous, and we had access to the \acrshort{refaap} results that the participants had completed.

\paragraph{Demographics}
Of 14 course participants, seven answered the survey. Three work as DevOps engineers, two as developers, and one in process quality. Two participants reported working in roles related to security.  
Two participants stated that security was ‘critical’ for their project. Two rated it as ‘important’, one as ‘somewhat important’, and two characterised security as ‘not important’.
Four of the participants work in a project team with fewer than twenty members, and three work in a project with more than one hundred stakeholders.

\paragraph{Perception about \acrshort{refaap}} Except for \textit{Simplicity}, \textit{Knowledge background}, and \textit{Overall attitude} the practitioners’ median rating for \acrshort{refaap} was ’7’, on a scale of ‘1’ to ‘8’, reflecting the overall usefulness of \acrshort{refaap}’s results for their real industry projects.
\begin{itemize}[label=\textbullet, nosep, leftmargin=*, topsep=4pt, partopsep=0pt]
    \item \textit{Simplicity: The self-assessment questionnaire was easy to apply}.  The practitioners rated the \acrshort{refaap} a ‘6’ out of ‘8’. One participant reported that some questions in \acrshort{refaap} were too vague and requested more concrete steps, while also acknowledging that vagueness is a common problem for most security assessment guidelines as they tend to be somewhat technology- and process-agnostic.
    \item \textit{Knowledge background: The assessment can be performed by stakeholders from departments that are not security specific}. The practitioners somewhat disagreed, rating this statement as ‘4’ out of ‘8’. The participants feared that non-security stakeholders might overlook important security actions due to inexperience. Surprisingly, participants who rated security as ‘critical’ or ‘somewhat important’ were least concerned by assessments performed by non-security stakeholders.
    \item \textit{Overall Attitude}. To discover their general opinion of \acrshort{refaap}, we asked the participants whether, given their security process improvement needs, they found \acrshort{refaap} better suited to their work than other approaches. Some participants reported being unaware of any alternative approaches.
\end{itemize}
\begin{table}[hpbt!]
\centering
\footnotesize
\begin{tabular}{|l|ccccccc|cc|}
\hline 
\textbf{Metric}& \textbf{P1} &\textbf{P2}  & \textbf{P3} & \textbf{P4} & \textbf{P5} & \textbf{P6} & \textbf{P7} & \textbf{Mean} & \textbf{Median} \\ \hline
Structuredness       & 7  & 6  & 2  & 7  & 7  & 7  & 7  & 6.1  & 7      \\
Simplicity           & 7  & 6  & 2  & 6  & 6  & 7  & 5  & 5.7  & 6      \\
Understandability    & 7  & 7  & 3  & 7  & 6  & 7  & 8  & 6.4  & 7      \\
Goal orientation     & 7  & 7  & 5  & 7  & 7  & 7  & 6  & 6.6  & 7      \\
Completeness        & 7  & 6  & 3  & 7  & 8  & 7  & 8  & 6.6  & 7      \\
Focus                & 7  & 7  & 6  & 7  & 7  & 7  & 8  & 7.0  & 7      \\
Sustainability       & 7  & 7  & 4  & 7  & 7  & 7  & 5  & 6.3  & 7      \\
Effectivity          & 7  & 3  & 4  & 7  & 8  & 7  & 8  & 6.3  & 7      \\
Efficiency          & 7  & 6  & 7  & 7  & 7  & 7  & 6  & 6.7  & 7      \\
Knowledge transfer   & 7  & 5  & 7  & 7  & 7  & 7  & 7  & 6.7  & 7      \\
Knowledge background & 7  & 4  & 4  & 3  & 5  & 7  & 1  & 4.4  & 4      \\
Overall attitude     & 7  & 5  & 6  & 7  & 5  & 7  & 5  & 6.0  & 6      \\
Overall suitability  & 7  & 7  & 5  & 7  & 7  & 7  & 8  & 6.9  & 7      \\
\hline
\end{tabular}
\caption{Study 8 Industry Course: Results of the survey regarding 
\acrshort{refaap}}
\label{tab:eval_study_2_assessment}
\end{table}
\greybox{Overall, the practitioners perceived \acrshort{refaap} as a structured approach that can be used to identify, introduce, and ensure security activities in their real-world projects.}

\paragraph{Usability of \acrshort{refaap} for industry} When asked whether they would follow the same procedure again if they had to assess the security of a DevOps project, seven out of eight practitioners responded affirmative. However, two participants mentioned that they would use it in conjunction with another framework and requested a complementary framework for more specific use cases, such as data protection and governance.

\greybox{Seven of eight practitioners would use the \acrshort{refaap} again if they had to assess security in a DevOps project.}

\paragraph{Perception about \acrshort{refa}} 
The median rating for different aspects of \acrshort{refa} was consistently ‘6’ out of ‘8’ or higher, thus highlighting its usability and value for industry use cases. Practitioners liked the \textit{\enquote{parallelism with SDLC [Software Development Life-Cycle] and DevOps phases}}. They also thought that it \textit{\enquote{can serve as a reference whenever there is doubt}} and that \textit{\enquote{it is applicable to most projects}}. The participants appreciated the additional documentation offered with \acrshort{refa}. However, one participant wished for a more extensive elaboration of the planning phase of \acrshort{refa}. One participant highlighted \acrshort{refa}’s usefulness in practice by stating, \textit{\enquote{I would be able to use the content in real life as well.}}.

\greybox{Practitioners perceive \acrshort{refa} as a valuable framework for DevOps projects in the introduction and analysis of security activities.}

\begin{table}
\centering
\footnotesize
\begin{tabular}{|l|ccccccc|cc|}
\hline
\textbf{Metric} & \textbf{P1} & \textbf{P2} & \textbf{P3} & \textbf{P4} & \textbf{P5} & \textbf{P6} & \textbf{P7} & \textbf{Mean} & \textbf{Median} \\ \hline
Ease of use          & 7  & 6  & 6  & 6  & 6  & 7  & 8  & 6.6  & 6      \\
Knowledge transfer   & 7  & 6  & 2  & 6  & 6  & 7  & 5  & 6.6  & 6      \\
Structuredness      & 7  & 7  & 3  & 7  & 6  & 7  & 8  & 6.7  & 7      \\
Effectivity          & 7  & 7  & 5  & 7  & 7  & 7  & 6  & 6.6  & 7      \\
Flexibility          & 7  & 6  & 3  & 7  & 8  & 7  & 8  & 6.9  & 7      \\
Overall attitude     & 7  & 7  & 6  & 7  & 7  & 7  & 8  & 6.6  & 7      \\
\hline
\end{tabular}
\caption{Study 8 Industry Course: Results of the survey regarding \acrshort{refa}}
\label{tab:eval_study_2_refA}
\end{table}

\subsection{Industry Pilot Study (Study 9)}
\paragraph{Goal}
With Study 9, we explore whether an assessment with \acrshort{refaap} can be performed by non-security stakeholders, and, if the results \acrshort{refamr} and \acrshort{refair} are useful. To cluster results, we define the two questions. (A) Could project teams perform the assessment themselves? And (B) does performing the assessment provide value to the projects?

\paragraph{Results}
\paragraph{(A) Can project teams perform the assessment themselves?} The assessment process involved all relevant stakeholders, including project managers, architects, and developers. Each stakeholder answered an excerpt of \acrshort{refaaq} of \acrshort{refaap} (see example in \Cref{tab:assessmentQuestionnaire}) and scored each current security maturity status indicator in their competence field individually using the \acrshort{refamr} maturity scale (see \Cref{tab:maturityScale}). This allowed us to compare their maturity scores with those of the security professionals in the projects. The comparison was made by averaging the scores of all non-security professionals and the scores of security professionals. We then computed the mean absolute difference between the two scores; the higher the average absolute difference, the greater the divergence between the perspectives of the security and non-security professionals.

\begin{figure}[hbpt!]
    \centering
    \includegraphics[width=\textwidth]{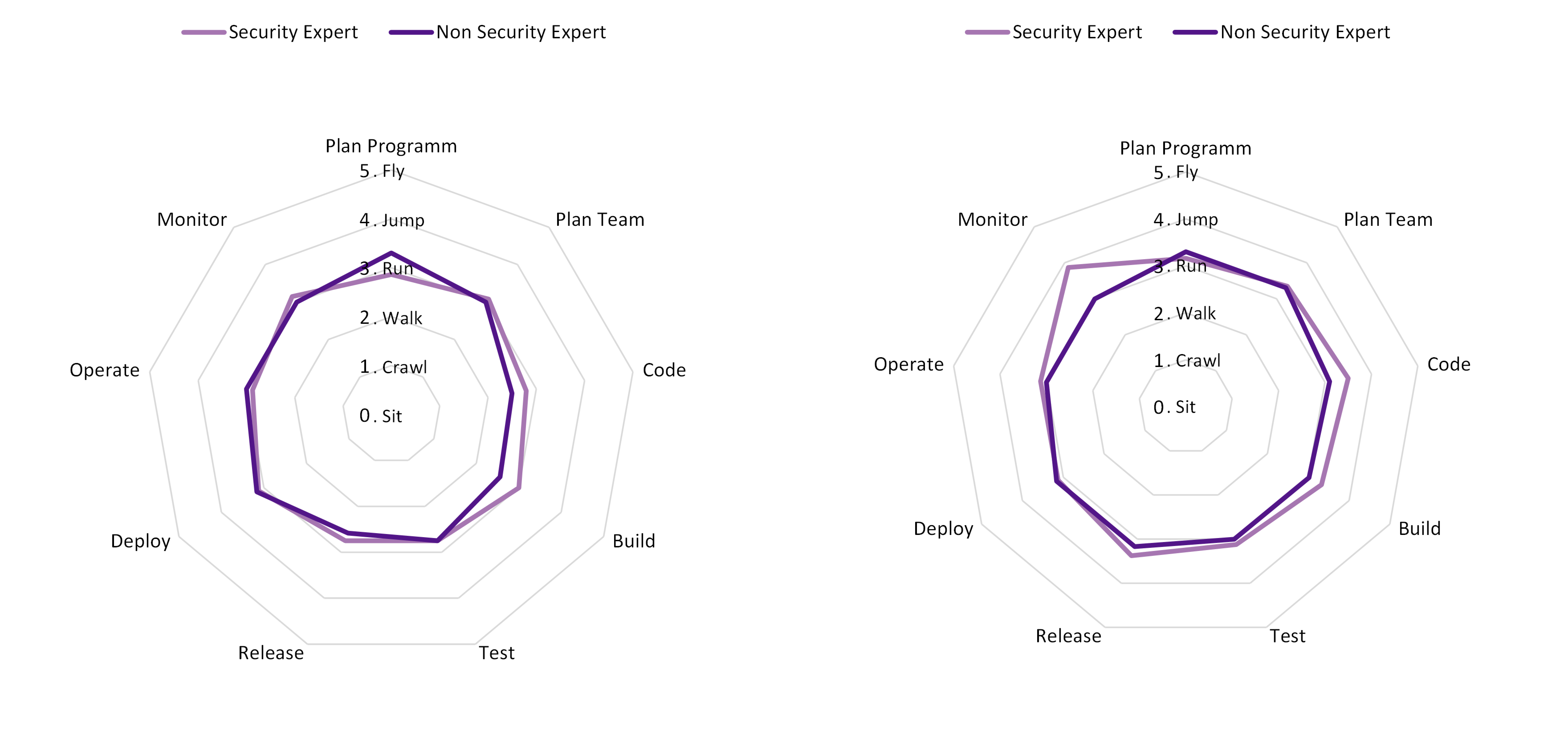}
    \caption{Study 9 Industry Pilot: Comparison of Mean Current Score (left) and Target Scores (right) of security practitioners and non-security practitioners assessments results. Scores are documented in \acrshort{refamr} following the scale presented in \Cref{tab:maturityScale}.}
    \label{fig:refa_sec_non_sec}
\end{figure}

The mean absolute score difference between security and non-security professionals ranged from 0.29 to 0.35 between the projects. These slight differences suggest that the teams could achieve similar results to those of security professionals. \Cref{fig:refa_sec_non_sec} depicts the mean maturity scores for security professionals and non-security professionals aggregated over all projects. For both perspectives, we note that the actual score and the target assessment score overlapped to a large extent.

During the three validation assessments, teams held in average 5 sessions and spent 6.25 hours in actual assessment work. This indicates that project teams can perform the assessment in a short timeframe. In addition, their perceptions about the security state were similar to those of security professionals. In conclusion, an assessment following \acrshort{refaap} can be completed within a sprint, allowing teams to plan the best timeframe with minimal or no dependencies on external teams, thereby enabling continuous assessment. We argue that the frequent application of the \acrshort{refaap} will positively impact the enablement of the security capabilities and support the overall agile security strategy of the organization. In addition, we note that the the reduced time investment on assessments impact developers efficiency and supports their focus in product development tasks.

\greybox{By applying our framework, engineering teams could perform security compliance assessments themselves in a short timeframe and with similar assessment scores as security professionals.}

\paragraph{(B) Does performing the assessment provide value to the projects?}
The assessment provided value for the practitioner's projects from three perspectives. First, the application of \acrshort{refaap} allowed stakeholders to openly discuss the current state of DevSecOps practices. The resulting \acrshort{refamr} provided a common shared representation of the current security state, that was easily understandable for all roles. It could be used in the broader continuous software engineering assessment process to leverage business decisions specifically targeting the organization's capabilities, for example, to more rapidly process customer feedback through continuous deployment and experimentation (c.f. \cite{olsson:2014:stairway}). Second, the project teams drew value from the aspects highlighted in \acrshort{refaar}. It enabled the improvmenet security of their development process based on \acrshort{refair} and directly integration into their project backlog. Examples of identified improvements included security training for secure coding and design, and shifting security activities further left in the DevOps lifecycle. Third, project managers perceived \acrshort{refaap} as helpful to determine the security status and want to further use it in conjunction with their existing security assessment approach.

\greybox{The assessment provided value from three perspectives: (i) visualization of current state of SC, (ii) discussion about target SC state, and (iii) decision making on required improvement actions}
\subsection{Final Expert Appraisal (Study 10)}
\paragraph{Goal}
This study incorporated an expert’s perspective to validate cohesion among the framework’s constructs. We expected that an expert in the assessment of industrial product development processes compliant with IEC 62443-4-1 would provide a critical view of the potential and limitations of our framework. The specific goals of this study were: (A) determine \acrshort{refa}’s \textit{accuracy} in representing security compliance practices and artefacts, (B) determine subjects perception about \acrshort{refa}, and (C) explore the \textit{applicability} of \acrshort{refaap} and \acrshort{refaar} for security compliance assessment in DevOps contexts.

\paragraph{Results}
The senior expert (henceforth, ‘the Subject’) has over twenty years of experience guiding engineering teams in implementing security processes with IEC 62443-4-1. The Subject is one of the initiators and keen supporter of the secure development framework used in the industrial context. The interview took place in 2023, and lasted 2 hours and 40 minutes. \acrshort{refa}, \acrshort{refaap}, and \acrshort{refaar} were made available for the Subject in digital format on several screens to facilitate his view, comparison, and exploration of the material. In the following, we summarize the results of the interview.

\paragraph{A. \acrshort{refa}'s accuracy in representing security compliance elements} 
The Subject was informed about the continuous development of \acrshort{refa}. We observed that the Subject needed little to no support to identified his focus points in \acrshort{refa}. As focus points, the Subject observed the following practice areas:code review, vulnerability identification in third-party components, project scope and applicability of security framework, vulnerability monitoring, and incident management. In addition, the Subject validated the presence of specific artefacts that should be included since they are relevant compliance evidence: secure coding standards, static code analysis report, documentation of security roles, statement of standard applicability, vulnerability reports, incident report, the definition of security roles and responsibilities, communication/escalation tree in case of incidents, and a list of trusted third parties to provide software components. The last two artefacts were missing in \acrshort{refa}. During the interview, we explained the rationale. However, for the Subject, the artefacts should be made explicit. \Cref{tab:pre_release_eva_focus_points_refa} presents a summary of the elements that were deemed relevant according to the Subject, together with the reference identifier in \acrshort{refa} if available. 

Besides their discussion of focus points, the Subject did not comment on the design elements of \acrshort{refa}. They seemed comfortable navigating the models. The Subject openly discussed and questioned the framework. For example: the relevance of the DevOps phase, where the security practice is located in \acrshort{refa}, the differences that exist between \acrshort{refa} and waterfall processes, or the maintenance complexity of the framework and the tools used for modeling. 
\greybox{The expert subject readily identified his focus points in \acrshort{refa}. He seemed confortable navigating the models.}
\begin{table}[hptb!]
\centering
\footnotesize
\begin{tabularx}{\textwidth}{|X|X|}
\hline
\multicolumn{1}{|c|}{\textbf{Subject's point of interest}} & \multicolumn{1}{c|}{\textbf{Matching RefA Practice with code and name}} \\ 
\hline
\textbf{Security code review} & \textbf{A-C9 Secure Coding Standards} \\
&\textbf{A-C8 Code Analysis}\\
- Coding standards & - C-13 Coding standards \\
- Code analysis report & - C-11 Peer-code review \\
  &- C-12 Static code analysis report \\ 
\hline
\textbf{Security roles and responsibilities}& \textbf{A-PP2 Team security responsibilities} \\
- Security roles & - PP9 Organization structure, roles and assigned personnel \\
\hline
\textbf{Selection of applicable security activities}&  \textbf{A-PP4 Security Standards Applicability Analysis} \\ 
- Applicable security practices&- PP15 Standard requirements applicable to the project \\
- Statement of standard applicability& - PP14 Analysis for standard's applicability to the project \\ 
\hline
\textbf{Security incidents management} & \textbf{A-M6 Security incident response} \\
- Incident report & - M10 Security incident report \\
- Communication/escalation channels & - \textit{Indirect match} to M11 Incident Plan \\ 
\hline
\textbf{Security of third-party components} & \textbf{A-C12 Third-party component selection} \\
- List of trusted third parties for software components & - \textit{No direct match} \\
\hline
\end{tabularx}
\caption{Study 10 Expert Appraisal: Matching of expert's focus points and the related practices in \acrshort{refa}}
\label{tab:pre_release_eva_focus_points_refa}
\end{table}
\paragraph{B. Expert's perception about \acrshort{refa}} 
The Subject rated the \acrshort{refa}’s over the same metrics as the participants in Study 8. \Cref{tab:pre_release_eva_refa} illustrates the ratings of the Subject, along with their remarks. This rating allows for a comparison to be made between the Subject’s perception of \acrshort{refa} and the perceptions of the engineering practitioners in Study 8 (c.f. \Cref{tab:eval_study_2_refA}). Similar to the engineering practitioners, the Subject found \acrshort{refa} easy to use, although they noted that this perception may have been influenced by prior knowledge of the security standard. They perceived \acrshort{refa}’s practice areas and artefact view to be well structured and effective in describing security compliance, at least with regard to the security activities that were their focus points.

The Subject appreciated the following \acrshort{refa} practice areas: \textit{continuous security testing} as it described the several types of testing and the findings treatments; \textit{continuous security feedback} as a novel area, and the implementation of the DevOps principles, continuous feedback and shift-left: \textit{``the message is clear, that finding should be solved as soon as possible''};  \textit{infrastructure code}, as it clearly indicates that the security lifecycle not only focuses on the application code but is adaptable to the product development project. This observation also influenced the Subject’s positive perception of \acrshort{refa}’s flexibility and adaptability for different types of DevOps projects. However, the Subject did criticise the framework’s flexibility with regard to its technical terms, including security activity, security practice, and security aspect.

\greybox{The Subject found \acrshort{refa} practice areas of \textit{continuous security testing}, and \textit{infrastructure code} as insightful. He found \textit{continuous security feedback} a novelty, and pointed out a missing artefact: list of trusted third-parties for software components.}

In contrast with the engineering practitioners, the Subject pointed out conflicts in the use of \acrshort{refa} to visualize security compliance practices (see his overall attitude rating of ‘4’ in \Cref{tab:pre_release_eva_refa}). They might instead use other known frameworks, such as the one they advocate for, the IEC 62443-4-1 standard and CMM+Secure. They also mentioned that the missing artefacts influenced their rating in this instance.
Finally, the Subject raised three concerns about implementing \acrshort{refa} in DevOps Projects: the difference in the framework’s implementation for organizations and for individual teams; stakeholders, and the setup to define when to roll out the whole model or only parts of it.

\begin{table}[hptb!]
\centering
\footnotesize
\begin{tabularx}{\textwidth}{|m{2cm}|>{\centering\arraybackslash}m{1cm}|X|}
\hline
\textbf{Metric} & \textbf{Rate}  &\textbf{Quotes}  \\ \hline
Ease of use          & 6  &  \textit{\enquote{Because I know IEC 62443-4-1, identifying artefacts like coding standards is easy}} \newline \textit{\enquote{I haven't seen this security feedback-loop before.}} \\ \hline
Knowledge transfer   & 7  & \textit{\enquote{The difference between application code and infrastructure code is clear}} \newline \textit{\enquote{The message is you should do it ASAP}}; referring to security practices and shift-left approach.\\ \hline
Structuredness       & 6 or 7  & \textit{\enquote{I haven't seen all, only one part. So a 6 or 7}} \newline \textit{\enquote{It takes the structure of the IEC 62443-4-1}}  \\ \hline
Effectivity          & 6  & \textit{\enquote{What are security practices? I like the term practice.}} \newline \textit{\enquote{Use the term practices consistently.} } \\ \hline
Flexibility          & 6  & \textit{\enquote{Keep consistency with the term security practice, instead of security aspect.}} \\ \hline
Overall attitude     & 4  & \textit{\enquote{I know three approaches to visualize security practices: a)Internal Framework, b) IEC 62443-4-1, c)CMM+Secure-based}} \newline 
\textit{\enquote{These artefacts are missing: Preferred suppliers list although the area is there, Project Business Impact in Plan, Incident Communication Channel in Incidents area.}} \\
\hline
\end{tabularx}
\caption{Study 10 Expert Appraisal: Results of the interview concerning \acrshort{refa}}
\label{tab:pre_release_eva_refa} 
\end{table}

\paragraph{C. Applicability of the assessment utilities for security compliance assessment}
The Subject evaluated \acrshort{refaap} and \acrshort{refaar} in their role as security assessor. The Subject focused on the same points as in the first part of the interview. Based on \acrshort{refa}’s practices and artefacts, they analysed the assessment questionnaire in \acrshort{refaap}, as well as the assessors’ guide (see \Cref{tab:assessmentQuestionnaire}). Subsequently, the Subject proposed several example answers and rehearsed potential ratings in the maturity scale provided in \acrshort{refaar} (see \Cref{tab:maturityScale}).
 
In general, the Subject found \acrshort{refaap}’s assessment questions to be well-structured and consistent. The same opinion was expressed for \acrshort{refaap}’s assessor’s guide, where the Subject identified key differences in the focus of continuous processes and linear processes. The Subject examined the novel areas of the framework in depth, including the \textit{continuous security feedback loop} and \textit{continuous security testing}. They remarked that the guide for the different testing types was useful and necessary for the assessment.

The Subject perceived \acrshort{refaar}’s improvement roadmap as useful and applicable for findings prioritization. They observed the improvement roadmap in a previous study (see Study 6 in \Cref{subsec:solution_construction}). During the interview, they discussed the origin of the quadrants, the typical overload associated with listing improvement actions, and how this could be reduced with \acrshort{refaar}. We argue that, in this context, the assessors’ skills in summarising findings play a role, and together with the maturity report \acrshort{refamr}, our deliverables remain focused on critical practices while allowing the reporting structure to generate alternative report formats, such as a presentation or text.

In contrast to \acrshort{refaar}’s improvement roadmap, the Subject was first introduced to maturity report \acrshort{refamr}, during the interview. They questioned whether a new maturity scale was required when assessing DevOps for security compliance. We mentioning the following observations in support for the maturity scale: (i) the learning outcomes during the solution construction phase (see \Cref{subsec:solution_construction}); (ii) the conclusions of researchers who demonstrated the need for maturity assessments in agile development \cite{Tuncel:2020:comparison, Tuncel:2021:scope}; and (iii) the origins of \acrshort{refamr} where the researchers’ maturity model was adopted \cite{Tuncel:2021:proposal} in conjunction with discussions with experts on the piloting of the framework in our industrial context. The Subject considered these reasons as valid support for the maturity scale, mentioning, \textit{``If there are already four years of research on it, then it is necessary''}. Later, the interview focused on the differences between \acrshort{refamr} and other existing maturity models, including one used for secure development assessments in the Subject’s company. The primary differences between the models lie in how maturity levels are addressed and their emphasis on DevOps principles.

The Subject proposed several examples and rated them based on \acrshort{refamr}’s explanation. The following examples were of interest to the Subject’s: (i) the organization defined secure coding standards, while teams perform manual code reviews and automatic checks; (ii) externals identify vulnerabilities during the product operation, and the organization has an incident response plan; (iii) developers freely select third-party components for the product. In general, the Subject approved of the explanations provided for each level and their ratings. They valued \acrshort{refamr}’s emphasise on DevOps principles such as automation, continuous everything, and team collaboration.

At the end of the interview, we obtained the Subject’s opinion via a survey. The responses and remarks are listed in \Cref{tab:pre_release_eva_ap}. Our deliverables are understandable, goal-oriented, and suitably focused on agile and DevOps processes. They believed that teams would struggle to define a target level for each practice and, based on their experience, observed that teams often require external help for assessments. 

The Subject’s overall attitude towards using \acrshort{refaap} and \acrshort{refaar} for security assessment in DevOps remained inconclusive. The primary concern was the divergence between our framework and OWASP-SAMM. We summarized the benefits of our assessment framework as its full integration within an agile/DevOps maturity assessment workflow, and our artefact-centric view that enables continuous assessment by product engineering teams. We then presented to the Subject documentation about other frameworks' missing focus in agile/DevOps workflows, as well as, anonymized results of the industry pilot study (Study 9). The Subject analysed these results and pointed out that the results were consistent with their knowledge in the industrial context. 
\begin{table}[hptb!]
\centering
\footnotesize
\begin{tabularx}{\textwidth}{|m{3cm}|>{\centering\arraybackslash}m{1cm}|X|}
\hline
\textbf{Metric} & \textbf{Rate}  &\textbf{Quotes}  \\ 
\hline
Simplicity         & 3  & \textit{\enquote{It is hard for assessees (engineers) to determine the target maturity level.}}  \\ 
\hline
Structure          & 3  & \textit{\enquote{Because of the target rating, leave only the current status.}}  \\ 
\hline
Understandability  & 6  & \textit{\enquote{Assessor's guide contains the look for}}  
\newline \textit{\enquote{Clear connection to compliance artefacts for code review}}\\ \hline
Goal Orientation   & 6  & \textit{\enquote{I agree, this provides results for DevOps}}  \\ \hline
Focus              & 6 or 4 & \textit{\enquote{It's useful for waterfall processes too.}}  \\ \hline
Sustainability     & 3  & \textit{\enquote{Maintaining questionnaires requires automation of utilities}} \\ \hline
Suitability for self-assessment    & 4  & \textit{\enquote{Teams always require help from others.}} \\ \hline
Overall attitude   & 4  &  \textit{\enquote{I am not sure how different is from OWASP-SAMM.}}  \\ \hline
\end{tabularx}
\caption{Study 10 Expert's Appraisal: Results of the interview on \acrshort{refaap} and \acrshort{refaar}}
\label{tab:pre_release_eva_ap} 
\end{table}

\subsection{Summary of Results per Research Question}
In the following, we summarize the results to answer the research questions.
\paragraph{RQ1: \rqone} 
\acrshort{refa} appears to reflect the artefacts ecosystem of DevOps and its security compliance. The participants of Study 8 described \acrshort{refa} as providing a comprehensive overview of the activities and stages in DevOps with regard to security compliance. They appreciated the clear distinction between security and non-security activities,  thereby facilitating security assessment of DevOps projects. The participants especially valued the structured nature of \acrshort{refa}, which rapidly facilitated insights into the current security state in their DevOps projects.
The main improvement was two missing activities that should be made explicit in the \acrshort{refa} (from Study 10). These findings suggest that the \acrshort{refa} adequately captures the relevant concepts of the DevOps ecosystem and security compliance, but that \acrshort{refa}'s comprehensiveness may benefit from further refinement in certain areas.

\paragraph{RQ2: \rqtwo} 
We conclude that \acrshort{refaap} can be used to assess the security compliance of DevOps projects. This claim was validated by seven participants in Study 8 and by three DevOps teams who successfully applied \acrshort{refaap} in Study 9. The participants stated that the structuredness, ease of understanding, and goal orientation of \acrshort{refaap} enabled its successful application. In Study 8, the participants expressed concerns that non-security professionals might miss relevant aspects in an assessment. The subject expert in Study 10 also stated that \textit{\enquote{Teams always require help from others}}. These observations suggest team members should conduct security assessments with a security expert. However, in Study 9, non-security professionals performed the assessment using \acrshort{refaap} with similar results to those of security professionals, highlighting \acrshort{refaap}’s overall suitability for continuous security compliance assessment by development teams themselves. Nevertheless, the participants were inconclusive about the simplicity of \acrshort{refaap}. The participants in Study 8 perceived \acrshort{refaap} as simple, even though they would have liked additional documentation. The subject expert in Study 10 recommended simplifications in determining the target maturity level in \acrshort{refaap}.

\paragraph{RQ3: \rqthree }   
\acrshort{refaar} offers useful results for stakeholders. \acrshort{refamr} makes the current status of the security practices in a project explicit and highlights potential risks. For instance, \acrshort{refamr} highlights the risk of undetected vulnerable third-party components when a software component analysis is not performed. The majority (seven out of eight) of the participants in Study 8 stated they would use \acrshort{refaap} and its reports in their real-world projects. Of these seven, two participants reported they would use it in conjunction with another framework to fit their use case better, or they would cherry-pick the most relevant aspects of \acrshort{refaap}. The willingness of the participants in Study 9 to incorporate \acrshort{refaap} into their existing assessment workflow underscores the value that \acrshort{refaar} can provide them.

In conclusion, based on the positive feedback from the participants, we assert that \acrshort{refaar} yields positive results. However, the participants’ feedback to combine \acrshort{refaap} with other frameworks suggests the need for a more holistic assessment.

%------------------------------------------------------------
\section{Key Results} \label{sec:results}
%------------------------------------------------------------
We have presented the development and evaluation of a framework that enables the assessment of security compliance in DevOps processes. The framework is a response to prevailing challenges in the IT/OT industry. During the research process, we gained several insights and drew several conclusions. Those are summarised below:
\begin{itemize}
    \item [R1]\textbf{Compliance with security standards is a relevant organisational objective that aligns well with continuous software engineering}. Continuous software engineering implies process improvement \cite{olsson:2014:stairway}, which is the ultimate result of security compliance. Security compliance perspectives benefit from continuous practices by (i) \textit{achieving compliance} through iterative and incremental implementation of security standard requirements, and (ii) \textit{assessing compliance} with a continuous and systematic analysis of gaps in the implementation of security standard requirements.
    \item [R2]\textbf{Continuous assessment of security compliance is methodologically feasible} and, as we argue, it is technically possible with automation (see our ongoing research in \cite{angermeir:2024:acsc}). With the application of \acrshort{refa}-based assessment utilities, \acrshort{refaap} and \acrshort{refamr}, we learned that security compliance assessments: (i) can be performed more frequently than regular yearly/quarterly approaches, (ii) can fit within a two-week iterations, (iii) can produce incremental results either per DevOps phase or \acrshort{refa} practice area and (iv) can be performed by DevOps team members with limited support from security experts, if they have the necessary tools. The following observations support these points:
    \begin{itemize}[label=\textbullet]
        \item The modularity of \acrshort{refa} and \acrshort{refaap} allows assessments to be tailored to the assessment’s scope and focus.
        \item Assessments applying \acrshort{refaap} and \acrshort{refaar} can identify security compliance gaps, delivering compliance roadmaps enabling teams to define priorities and assess their impact. Our framework provides adequate results even when used by non-security practitioners (see \Cref{fig:refa_sec_non_sec}).
        \item Compliance assessments can be established in the workflow of agile and DevOps teams. The key success factors for such establishment are: (i) agile/DeOps team members should perceive achieving compliance as progress, (ii) the assessment environment should be open enough to discuss failure, (iii) the team should be flexible enough to integrate the opinions of stakeholders who have an interest in improving product’s security compliance, (iv) assessors, external to teams, should embrace agility and DevOps values and principles.
    \end{itemize}
    \item [R3]\textbf{The continuous implementation of security compliance is methodologically and technically feasible}, especially with the automation of security activities in CI/CD pipelines. During the introduction of \acrshort{refa} and the piloting of \acrshort{refamr}, practitioners described roadmaps to implement prioritised security practices. The following set of practices should be implemented incrementally and, represent a \textit{minimally viable} product for security compliance in DevOps: (A)~Plan: an analysis of the applicability of security standards and threat modeling to define product security requirements and secure design. (B)~Code: a static analysis of code and integration of secure coding standards. (C)~Build: a static analysis of code in CI, a vulnerability analysis of third-party software based on a software bill of materials, inspection of hardening policies implementation, and ensure integrity of distribution files. (D)~Test: dynamic application security testing, run-time testing, smoke testing and continuous security findings management. (E)~Release: distribution files integrity verification and security guidelines preparation. (F)~Deploy: provenance assurance and secrets management. (G)~Operate: secure cloud services configuration, cloud identity and access management, and security updates execution. (H)~Monitor: continuous monitoring of the application and environment, including security incident response. 
   \item[R4] \textbf{There are gaps between achieving compliance with the industrial security process standard, and achieving secure DevOps}. The IEC 62443-4-1 standard does not fully cover the DevOps lifecycle, as it leaves the operational phases (Ops) uncovered. Companies should rely on other process standards or technological best practices to complement security in Ops. For example, the IEC 62443 sibling, \textit{ 2-4 Security program requirements for IACS service providers} \cite{iec:24:2023}, which covers secure operation, or the \textit{NIST SP 800-218 Secure Software Development Framework} \cite{nist:2022:ssdf} that can be used to analyse requirements for release integrity, or the \textit{Cloud Control Matrix} from the Cloud Security Alliance \cite{csa:2024:v4} to cover security operations practices in cloud deployments. The IEC 62443-4-1 standard requires specific interpretations to cover continuous integration and delivery (CI/CD) for the Dev phases. For example, the secure implementation practice focuses on application code, but disregards infrastructure or pipeline code, and does not explicitly describe artefact provenance and inventory (SBOM). In our view, \acrshort{refa} adequately covers the security practices in a DevOps lifecycle, including the IEC 62443-4-1 practices and others identified during the artefacts analysis in our literature review and our observations of continuous security compliance in practice.
\end{itemize}

%------------------------------------------------------------
\section{Limitations and Threats to Validity} \label{sec:validity}
%------------------------------------------------------------
This research is subject to several limitations. In this section, we focus on three of them. First, the context in which we worked limited data sharing through non-disclosure agreements. We mitigated this threat by describing the studies in as much detail as possible, sharing quotes or aggregated data after prior approval. Second, we avoided selection bias by entrusting independent moderators to invite and select participants for the study. Existing expert communities were also used for validation. Third, the primary author’s role in the large company raised concerns about confirmation bias, which were mitigated by implementing protocol tracking among several research partners, including the co-authors. The following sections discuss the measures taken to address internal, construct, and external validity threats in our studies (see \Cref{fig:method}).

\paragraph{Solution Construction: Studies 3 to 7}
Confirmation bias was a significant concern regarding the internal validity of our studies. Confirmation bias was reduced by involving several researchers in each study and performing preliminary evaluations with practitioners in controlled environments, for example, where at least one researcher focused on the study’s rigour. We are also aware that frequent preliminary evaluations may raise an external validity threat about the representativeness of the stakeholders. This threat was triggered since \acrshort{refa}, \acrshort{refaap}, and \acrshort{refaar} were pre-evaluated by individuals from the same organisation. It remains unclear whether this study’s findings and contributions would have been the same if the construction process had been performed in a different organisational context. To counter external validity threats, we extensively evaluated the framework (Studies 8-10) involving other organisations.

\paragraph{Solution Evaluation: Studies 8 to 10}
Since security compliance and DevOps constitute a still-maturing field of study, the limited expertise of some participants may have influenced the results of our studies, for example, by triggering false assumptions. We mitigated this internal validity threat by evaluating our constructs across three studies in different organisations, this helps to mitigate the threat of trust on the opinion of one expert (Study 10), who is not part of the research team, rather a driver of the existing assessment frameworks. Threats to the construct validity are twofold. First, we prevented participants from misunderstanding the framework by providing them with training (Study 8) or paced guidance (Studies 9 and 10) before they used it. Second, we avoided (unintentional) data poisoning and minimised interpretation bias by using a Likert scale from ‘1’ to ‘8’ supported by verbal and written explanations. Researchers outside the research group aggregated the results of these studies. Finally, we address threats to the studies’ generalisability by deriving accurate indicators for each research question and ensuring that each question is addressed in at least two studies.

%------------------------------------------------------------
\section{Conclusion and Future Work} \label{sec:conclusion}
%------------------------------------------------------------

In this paper, we present a problem-driven longitudinal research program entailing a series of empirical studies (see \Cref{fig:method}), involving 102 practitioners who have participated in the development process of twenty-six industry software products, twelve of which are related to our industrial context. These studies enabled the identification of relevant challenges and needs in the field of security compliance (SC) within agile development and DevOps. We framed the requirements and evaluation stage of our framework to enable SC assessments of DevOps processes, resulting in the \textit{reference architecture for security-compliant DevOps} \acrshort{refa} and the related \textit{assessments utilities} \acrshort{refaap} and \acrshort{refaar}. \acrshort{refa} embeds an ideal security-compliant DevOps process by mapping practice areas and artefacts that are security-relevant according to the IEC 62443-4-1 standard requirements, state-of-the-art literature, and industry practices. \acrshort{refaap} supports the systematic SC assessments for DevOps processes, and \acrshort{refaar} visually compiles SC results that are clear and understandable to security practitioners and stakeholders in business, development, and operations.

As a result of the challenges investigated during the problem analysis of this research program (see Studies 1 and 2  in \Cref{subsec:problem_analysis}), we have learned that industry practices are undergoing a methodological shift from traditional processes with decisive security gates to fluid process flows. This new approach empowers DevOps engineering teams to make informed decisions regarding secure product development and functional security with reduced reliance on security experts, a scarce and costly resource. Organisations are increasingly interested in incorporating security knowledge into engineering teams and utilising automated support for security activities. For this to be effective, engineers require transparency in the security status of development processes and the product itself. We built \acrshort{refa} and the related \textit{assessment utilities} to allow such transparency at the process level.

During our framework's construction and continuous improvement (see Studies 3 to 7 in \Cref{subsec:solution_construction}), we observed that security assessment frameworks relying on activity-based models transfer limited security knowledge to engineering teams. However, they do support other stakeholders’ understanding of security issues. To improve security compliance in DevOps processes, the engineering team must thoroughly understand the details of the security practices if they are to implement them efficiently. We provide information about security compliance in \textit{\acrshort{refa}'s artefact-view}, where security artefacts with DevOps artefacts are presented. We also offer \textit{\acrshort{refa}'s areas view}, which is aimed at stakeholders who require a high-level analysis (see \Cref{fig:dso_build}). 

The evaluation of our framework (see Studies 8 to 10 in \Cref{sec:evaluation}) supports our contribution as an instrument for engineering teams to visualise and track the security status of their DevOps processes, providing transparency and supporting decision-making for stakeholders. 
Applying \acrshort{refa}-based assessment utilities: \acrshort{refaap} and \acrshort{refaar}, practitioners observed that our framework enhances the competence of engineering teams, as they can perform SC assessments more frequently and are less dependent on security compliance assessors. We have learned that when engineering teams are aligned with the DevOps transformation, they should conduct continuous SC assessments in accordance with agile principles. Achieving compliance is perceived as progress, and the assessment settings are flexible enough to fit with iterations. We recommend that SC assessments be registered as backlog items, according to a schedule, thereby keeping team members focused on producing working software. We also recommend that SC assessors act like agile coaches, guiding team members to identify relevant gaps while approaching non-compliant security practices or existing vulnerabilities with the same constructive attitude that agile practitioners adopt toward failure.

Currently, \acrshort{refa} is used by practitioners in several companies as a means to gain an understanding of security practices in DevOps. A large industry partner currently utilises it to enhance the consistency of its existing reference models with continuous security practices and to elicit security capabilities for platform engineering systems. Meanwhile, the \acrshort{refaap} and \acrshort{refaar} assessment utilities are employed partially as modules within existing assessment methodologies or as input to refine and improve these methodologies. This partner has provided valuable feedback that will be used for subsequent versions of our framework.

With our framework, we envision practitioners acquiring or enhancing their knowledge of security in DevOps, promoting continuous assessments and forming cross-functional teams that possess the skills required for SC implementation and assessment. In this context, the reference architecture \acrshort{refa} constitutes a single framework that supports stakeholders from business, engineering, and security to map security standard practices in their processes. Based on the observations from our studies, we believe that using a single framework will increase awareness and collaboration, thereby enhancing CSC improvement. Practitioners may use the \textit{minimum set} of practices for security-compliant DevOps as a guide for implementation (see result R3 in \Cref{sec:results}).

Researchers can utilise our framework to visualise the integration of security practices into DevOps phases, enabling them to identify areas for improvement in both their application and deployment. In addition, investigators seeking to integrate other security- or quality-related standards, such as reliability and safety, can replicate the methods used to develop this framework.

During our studies, practitioners highlighted differences between our framework and OWASP-SAMM. We, together with several industry practitioners determined the scope and focus of such assessment framework to be limited in its coverage of product development security in continuous software engineering contexts. While it enables organisations to report on the maturity of their security programs, it leaves product development aspects uncovered, for example, artefacts.

\acrshort{refa} has boundaries. While we do not claim that \acrshort{refa} is a one-size-fits-all model, our evaluation results give us confidence that it is suitable for continuous software engineering, especially compared to existing approaches. Practitioners rated the integration of continuous software engineering and secure software engineering practices in DevOps lifecycles as understandable, well-structured, and effective for knowledge transfer purposes. 
However, we identified the following possible enhancements: (i) Specific versions of \acrshort{refa} may be required to illustrate security risk management throughout a product’s lifecycle, for example, to explicitly classify security findings as risk-based in each of the DevOps phases and later integrate them into the backlog as items. (ii) \acrshort{refa} requires an extension for specific security aspects of the architecture used in cloud-native environments that cannot be captured in a regular process assessment. (iii) \acrshort{refa} may be improved to cover implications on security practices caused by changes in collaboration structures between DevOps teams that are not well-integrated. In our experience, barriers between development and operation teams remain; a topic of relevance for future research.  
Regarding the assessment utilities, one constraint is that \acrshort{refamr} shows the average for \acrshort{refaap} responses. A more in-depth statistical analysis may reveal whether responses to specific questions should have a greater impact on the maturity score.

In the future, we will improve our deliverables by drawing on the lessons learned from our industry use-case study. We also plan to provide automated tools to facilitate the maintenance of \acrshort{refa}, for example, by automatically generating models based on digital catalogues of practices and artefacts. This will support use cases of scaling security process assessment across several teams while allowing practitioners to collect and aggregate assessment results in a single pool. Finally, we plan to implement systematic mappings of \acrshort{refa} artefacts with other security norms and frameworks used by the industry, including the NIST SP 800-218 Secure Software Development Framework. 
%------------------------------------------------------------
\section{Data Availability} \label{sec:data}
%------------------------------------------------------------
\acrshort{refa} is available at \cite{moyon:2023:refa}. The survey instruments used in the studies reported in this paper, as well as excerpts of our assessment utilities (\acrshort{refaap} and \acrshort{refaar}), are publicly available as open materials under \cite{this_online_material}. However, the raw data collected during our studies cannot be disclosed due to confidentiality agreements with our industry partners. 
%------------------------------------------------------------
%------------------------------------------------------------
\section{Acknowledgements} \label{sec:acknowledgements}
%------------------------------------------------------------
We are grateful to our anonymous reviewers and colleagues, especially those in the SAG-SAD task force, whose continuous feedback improved our approach and its practical application. We appreciate the junior researchers whose work planted the seeds for this compiled study: Svitlana Ierastova, Pranabi Bitra, and Shrileka Achanta. Finally, we are grateful to our participants, whose trust in the research process made this contribution to our understanding of the field possible.

\bibliographystyle{elsarticle-num-names}
\bibliography{main.bib}

%------------------------------------------------------------
\newpage
%------------------------------------------------------------
\section*{Appendix} \label{sec:appendix}
%------------------------------------------------------------
\subsection*{Terminology used in this Paper}
\Cref{tab:02_relevant_concepts} lists the foundational concepts and terminology used in this paper. 
{
\begin{table}
\scriptsize
\centering
\begin{tabularx}{\textwidth}{|l|X|}
\hline
\multicolumn{1}{|c|}{\textbf{Term}} & \multicolumn{1}{c|}{\textbf{Definition}} \\
\hline
\multicolumn{2}{|c|}{\textbf{Software Development}} \\
\hline
Continuous Software Engineering  (CSE) & Methodology that emphasizes iterative and incremental development of software products to improve product and service quality continuously \cite{Bosch:2014:continuous,Fitzgerald:2017:Continuous}. \\
\hline
Agile Software Development & A set of principles and practices based on the agile manifesto~\cite{beck:2001:agile}, embracing collaboration, frequent changes, customer involvement, early and frequent continuous delivery and consistent improvement \cite{beck:2001:manifesto}. \\
\hline
DevOps & \enquote{Set of principles and practices which enable better communication and collaboration between relevant stakeholders to specify, develop, and operate software and systems products and services, and continuous improvements in all aspects of the life cycle} \cite{iso:32675:2022}. DevOps is typically structured according to the following phases: Plan, Code, Build, Test, Release, Deploy, Operate, and Monitor. \\
\hline
Continuous Integration (CI) & A software development practice involving the frequent integration of code changes and testing them for defects. CI supports the concept of 'early feedback' by moving testing activities earlier in the DevOps pipeline, also known as 'shifting left'. \\
\hline
Continuous Delivery (CD) & A software development practice consisting of automating release and deployment procedures, based on continuous integration. \\
\hline 
Shift Left & A software development principle allowing the practitioner to perform activities in earlier steps in the engineering workflow, thereby allowing for fast feedback loops. \\
\hline
Early Feedback Concept & A software development principle that describes the integration of feedback loops during the early phaes of an engineering workflow. \\
\hline
\multicolumn{2}{|c|}{\textbf{Security and Compliance}} \\
\hline
DevSecOps & \enquote{The DevOps view of security is sometimes referred to as DevSecOps, but in reality, there is no DevOps without a continuous focus on security.}\cite{iso:32675:2022}. \\
\hline
Regulator & An entity that provides normative documents such as standards, laws, or policies, and may assesses a product's compliance with such documents. \\
\hline
Compliance & The adherence of a software product or service to requirements originating from normative documents such as standards, laws, or policies. \\
\hline
Compliance Assessment & The analysis of how a software product or service complies with requirements set out in normative documents.\\
\hline
Security Standard & Normative document prescribing rules and best practices for establishing and maintaining security. A security standard often focuses on product's security by design, security by default and secure operation. \\
\hline
IEC 62443-4-1 & This code refers to a certifiable security standard that defines the process requirements for the secure development of industrial automation and control systems. This standard is part of the widely recognized IEC 62443 family of standards \cite{iec:62443:2014}, and is widely used by industrial companies. \\
\hline
\multicolumn{2}{|c|}{\textbf{Research Methods}} \\
\hline
Delphi & A technique for structured communication with experts in expert panels to reach a consensus on a particular topic while navigating threats in typical expert panels. \\
\hline
Lean Coffee & A meeting format that allows participants to freely and collaboratively discuss topics they deem relevant without being limited by a predefined meeting agenda \cite{dalton:2018}. \\
\hline
\multicolumn{2}{|c|}{\textbf{Others}} \\ 
\hline
Maturity Model & A framework used to assess and improve the maturity level of a target, such as a particular development process. Such models are typically based on an existing norm. \\
\hline
Reference Architecture & A blueprint outlining guidelines regarding the implementation of a specific work architecture, for example, of a development process. \\
\hline
Quality Gates & Checkpoints in the development process that prevent the execution of further activities if a given set of quality requirements is not fulfilled. \\
\hline
Risk-based Approach & An approach used in identifying, evaluating, prioritizing, and treating potential risks. This approach allows for informed decision-making. \\
\hline
\end{tabularx}
\caption{Definitions of the terms used in this paper.} 
\label{tab:02_relevant_concepts}
\end{table}}

\subsection*{RefA Construction Summary}
\begin{figure}[htp!]
\centering
\includegraphics[width=\textwidth]{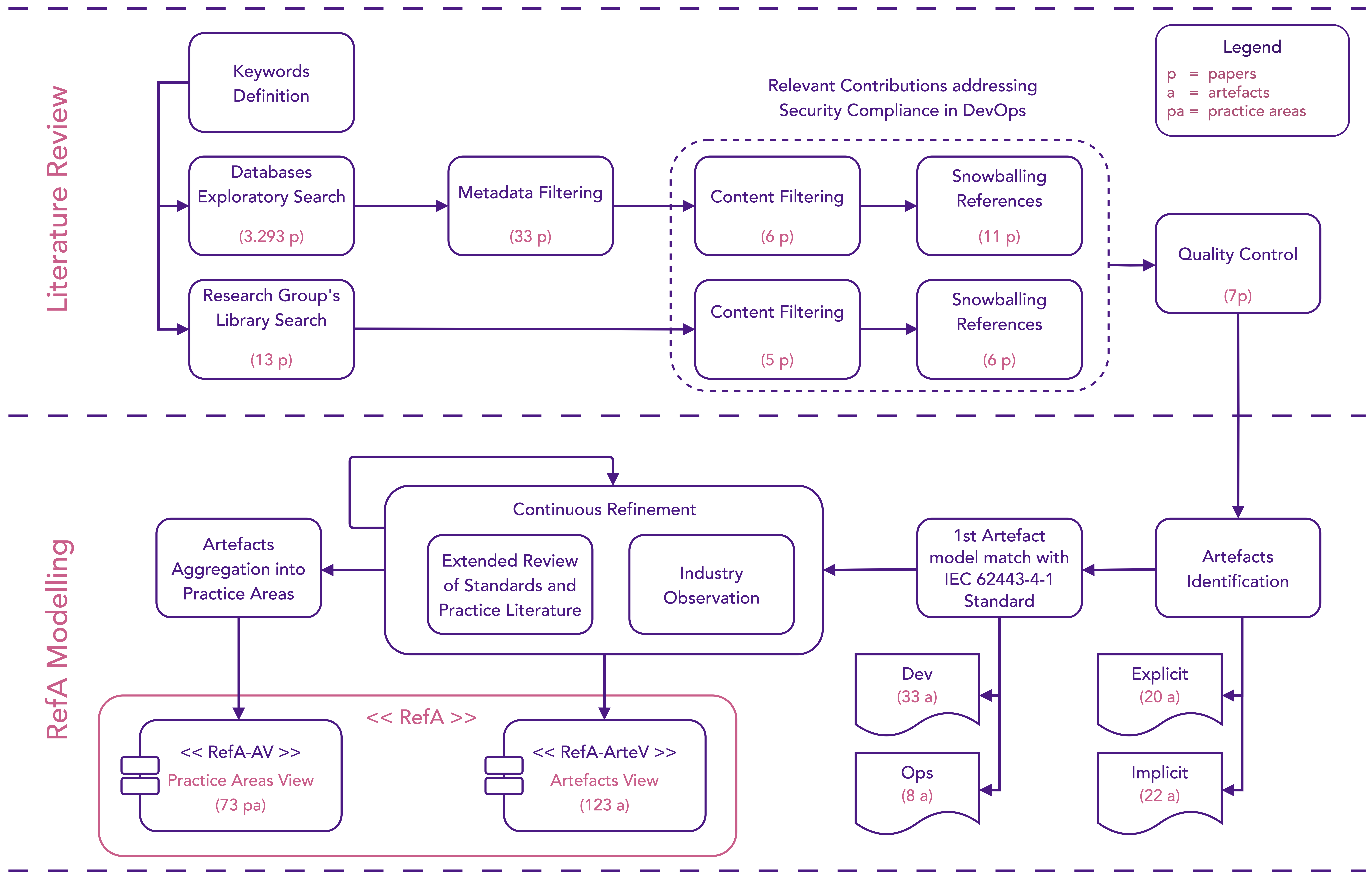}
\caption{An overview of the construction of \acrshort{refa}: Summary of procedures to develop the Reference Architecture for Security-Compliant DevOps. The steps regarding the 1st Artefact Model Proposal are adapted from \cite{Bitra:2021}.}
\label{fig:artefact_method}
\end{figure}

\subsection*{\acrshort{refa} Lifecycle}
\Cref{tab:refaLifecycle} summarizes per each DevOps phase the \acrshort{refa}'s security-compliant DevOps lifecycle noting its \textit{artefacts} and \textit{practices} in italics. The lifecycle begins by splitting the \textit{Plan phase} into two flows: (A) the flow where value is delivered through interacting teams, and (B) the flow where an individual team delivers features. \acrshort{refa} names these flows according the Scaled Agile Framework SAFe \cite{Leffingwell:2017:SAFE}, i.e., DevOps Plan at the \textit{Program}, and \textit{Team Levels}. \acrshort{refa} later describes the practices from the \textit{Code} phase to the \textit{Monitor} phases. 

{
\centering
   \begin{table}
    \centering
    \footnotesize
    \begin{tabularx}{\textwidth}{|X|} 
    \hline
    \textbf{Plan:} At the Program Level \acrshort{refa} represents a strong integration of security into the value stream. The \textit{program backlog} includes \textit{process-related requirements} such as managing team security training and capabilities, as well as an applicability analysis of security standards for the products in the value stream. At the Team Level, \acrshort{refa} proposes cross-functional teams to list security \textit{product-related} and \textit{process-related requirements} as backlog items. The elicitation of functional \textit{security requirements} may require specific \textit{technical security standards} for the \textit{operational environment} in which the software system will run. Security requirements shape the \textit{secure design} which is part of the \textit{defense in depth} approach. This involves addresing \textit{threats} iteratively and incrementally, guided by a \textit{security risk analysis}.\\
    \hline
    \textbf{Code:} Engineers implement \textit{product-related requirements} with application code and code to support the application's build, packaging, and delivery. This may include its infrastructure, CI/CD pipeline, or test code. Code should be aggregated in a \textit{development branch} that tracks its \textit{versions}. Teams should i) regularly \textit{analyze code} to detect and address vulnerabilities regularly automatically and/or manually, and ii) establish \textit{coding standards} to avoid vulnerabilities. \textit{Private keys} or other credentials pertain to {secrets vaults}. DevOps teams provisioning environments should automatically \textit{harden infrastructure resources}, for example, containers, as early as possible. Early in this phase, engineers can implement \textit{software composition analysis} techniques to generate a \textit{software bill of materials (SBOM)} and further detect \textit{vulnerabilities in third-party components}. DevOps principles of continuous feedback should apply to the security activities mentioned here.\\
    \hline
    \textbf{Build:} With continuous integration, engineers tune the pipeline to automatically analyse code for vulnerabilities and any changes in high-risk code. This pipeline tuning requires continuous learning and collaboration based on \textit{integration reports}. \textit{Supply chain security} is important to detect vulnerabilities in \textit{third-party components} listed in the \textit{SBOM}. Compliance with \textit{security hardening policies} should be checked for \textit{infrastructure resources}, while security functions require unit testing as part of the integration environment. In a continuous build, there should exist means to validate every \textit{distribution file’s origin and integrity}, for example, by using digital signatures. Finally, the identified issues are treated according to \textit{security findings management} techniques.\\
    \hline
    \textbf{Test:} \acrshort{refa} notes that a software-intensive system's application and run-time environment can be generated from a single \textit{distribution package}. Therefore, \textit{environments management} should be used to ensure the mirroring of testing environments (developer, staging, and user acceptance) to the production environment. Automatic security testing practices performed by the CI/CD pipeline or on demand, as well as manual security testing practices, are aggregated into \textit{continuous security testing}, for example, user acceptance testing, fuzzy testing, or vulnerability testing. These techniques apply to the application and the run-time environment. Finally, \textit{continuous treatment and feedback of security findings} practices utilise automatic processes to sort and organise test results continuously.\\
    \hline
    \textbf{Release:} With \textit{continuous delivery}, the \textit{signed} and \textit{tested distribution package} is forwarded as a \textit{release candidate} whose integrity, and availability should be ensured through its \textit{provenance} and \textit{Repository Protection}. In this phase, a \textit{software composition analysis (SCA)} is applied to deliver an accurate \textit{SBOM} to re-check vulnerabilities in components. Finally, \textit{security guidelines} should be prepared to explain security deployment instructions, including software configurations, accounts management, and network setup. \\
    \hline
    \textbf{Deploy:} \textit{Continuous delivery} and \textit{}\textit{continuous deployment} allow engineers to roll the \textit{release candidate} into the \textit{production environment}. The CI/CD pipeline validates the \textit{origin} of the \textit{distribution package}, and the \textit{deployment script} acts. RefA simplifies the deployed system into: \textit{application}, \textit{run-time environment}, and \textit{data storage}. The \textit{production repository} and \textit{deployment scripts} should be protected, as well as any credentials used. Therefore, a \textit{secrets vault} is needed. Finally, all \textit{security guidelines} prepared during the \textit{release phase} must be used during the deployment phase.\\
    \hline
    \textbf{Operate:} The \textit{production repository} and \textit{secrets} should be protected against unathorised access. Backup management practices should also be implemented. \textit{Network nodes} should be segmented even in \textit{cloud environments}. The related \textit{cloud services} should be hardened with \textit{security configurations} compliant with organization\textit{ policies}. Previous \textit{security controls} should be listed in the \textit{security guidelines}. Continuous security updates must be implemented to ensure timely remediation of security vulnerabilities \\
    \hline
    \textbf{Monitor:} Security monitoring is based on a \textit{continuous monitoring plan} that targets the \textit{running system: application, environment and data storage}, the \textit{production repository}, and the \textit{network such as cloud service}. \acrshort{refa} describes automatic and manual monitoring activities such as run-time checks, vulnerability scans, red/blue team exercises, and bug bounties. The \textit{continuous feedback practices} direct \textit{security findings} to the left phases. Teams should establish \textit{security incident plans} to ensure appropriate handling of security incidents. \\
    \hline
    \end{tabularx}
    \caption{Summary of the security-compliant DevOps lifecycle described in \acrshort{refa}. Italics denote artefacts and practice areas.}
    \label{tab:refaLifecycle}
\end{table}
}
\subsection*{RefA Assessment Questionnaire}
\Cref{tab:assessmentQuestionnaire} shows an excerpt of the \acrshort{refaaq} \textit{assessment questionnaire}. See \Cref{fig:refa_elements} for further information about this component of our framework.
\begin{table}[ht!]
\scriptsize
\centering
 \begin{tabularx}{\textwidth}{p{0.1\textwidth} p{0.26\textwidth} X}
  \toprule
  \multicolumn{1}{c}{\textbf{RefA's Area}} & \multicolumn{1}{c}{\textbf{Question}} & \multicolumn{1}{c}{\textbf{Assessor's Guide}} \\ 
  \hline
\multicolumn{3}{l}{\textbf{DevOps Stage: Code}}\\ \hline
{A-C1 Product Requirements Implementation} & {Is it possible to trace functional security requirements with the related code, for example, branches?}& {Understand how security functional requirements are developed. Determine if they are included as requirements and part of the same code repository as the rest of the application. Focus on traceability to validate whether the secure design has been implemented.} \\
\midrule
A-C4 Everything as Code & Is there a homogeneous development process for applications, infrastructure, pipelines, etc? & Understand if the flow to elicit, implement, and test deployment requirements for application functions is the same as the flow for infrastructure, pipelines, and any other requirements. If not, describe the differences and identify missing flows concerning security.  \\ 
\midrule
{A-C8 Code Analysis} & Are manual code review practices e.g. peer-review effective in improving security in code continuously? & Understand if manual code review practices are implemented to the point where findings are collected and shared among the team, with the objective of learning about mistakes and if specific security issues are treated. Determine if SAST tools are updated to avoid findings detected through manual reviews. \\
\toprule
\multicolumn{3}{l}{\textbf{DevOps Stage: Build}}\\ \hline
 {A-B3 Code Analysis and High-Risk Code Tracking} & {Are automatic code review practices effective in reducing security risk and continuously improving security? For example, SAST analysis with a security focus before the commit phase, or  incremental SAST for code at rest.}& {Understand how SAST Tools work during continuous integration and continuous build activities. The implementation strategy should guarantee that most of the code is reviewed incrementally. Regarding security, SAST analysis should effectively detect security issues. Ask if the SAST tool uses ad-hoc configuration or its default configuration. Enquire whether any tool provides information about changes in High-risk code. Determine if security findings have been analysed and solved, for example, findings identified by the SAST tool or changes in high-risk code.} \\ \midrule
 {A-B5 Software Bill of Materials} & {Are there mechanisms to automatically generate a Software Bill of Materials (SBOM)?}& {A SBOM is a comprehensive list of every component within a system including both OSS and COTS 3rd-Party Components. To continuously keep the Software Supply Chain working, it is essential to keep the SBOM updated. This allows us to identify necessary updates for the components early on, determine where to apply them, and communicate effectively with the client about known vulnerabilities in the product.} \\ 
    \midrule
{A-B7 Container Security Compliance Hardening} & {Are practices to validate compliance with recommended security configurations in containers or other similar technologies effective in continously reducint security risks (i.e. security compliance hardening) ?}& {Understand how the pipeline works in order to validate that container images, VMs, and OSs comply with recommended security configuration and security hardening policies defined either internally or by an approved external party. For example, CIS Benchmarks can be used as a reference for hardening OSs, and tools use these Benchmarks to validate compliance. Determine whether findings are addressed promptly, for example, are they aggregated as part of the security findings repository or trigger a backlog item for the respective team in charge.} \\ 
 \midrule
 {A-B10 Build Repository Protection} & {Are security controls implemented to protect artefacts produced during the build phase?
Are such controls documented as secure process requirements and part of the backlog of at least one team?}& {It is essential to understand how the build repository is protected, as well as how reports and artefacts, such as keys to sign distribution files, commit reports, build reports, third-party vulnerability reports, unit test report, and hardening policies, are managed. Determine who is responsible for each of these controls. Consider that, in large organizations, these tasks may be done by another team in charge of IT of the whole organization, while in Small to medium organizations this could be done by a specific team part of the Value Stream, which also develops functionalities, or by the Ops team.} \\ 
\toprule
\multicolumn{3}{l}{\textbf{DevOps Stage: Test}}\\ \hline
{A-T3 System in a Run-Time Testing Environment} & {Is the system in the test environment wholly and directly generated from the signed distribution package provided by the pipeline?}& {This question is especially relevant when infrastructure as code is used. The question implies that the pipeline produces a signed distribution package, which is then placed in a  test repository by a separate script or by the pipeline itself. Afterward, the script would activate services, including both the application and run-time environment (for example, a component running in a container), thereby bringing the system online} \\\midrule
{A-T4 Continuous Security Testing} & {Are practices to identify application security vulnerabilities through dynamic security testing (DAST) effective in continously reducing security risks continuously?}& {Dynamic Application Security testing enables the automatic identification of vulnerabilities in the application. Identify which tools are being used to perform DAST and certain whether they are being used and whether they are triggered automatically by a pipeline or manually, on demand?} 
\\
  \hline
\end{tabularx}
\caption{\acrshort{refaap}: Excerpt of the Assessment Questionnaire \acrshort{refaaq} for the DevOps Phases: Code, Build, and Test.}
\label{tab:assessmentQuestionnaire}
\end{table}
\subsection*{Maturity Scale to evaluate security practice areas in \acrshort{refaar}}
\Cref{tab:maturityScale} shows the maturity scale used to derive \acrshort{refamr} \textit{maturity report}. See \Cref{fig:refa_elements} for further information about this component of our framework.
{
\begin{table}[ht!]
\scriptsize
\centering
 \begin{tabularx}{\textwidth}{p{0.02\textwidth}X X X X p{0.21\textwidth} }
        \toprule
        \multirow{4}{=}{\rotatebox[origin=c]{90}{\textbf{Maturity Level\phantom{00}}}} & \multicolumn{5}{c}{\textbf{Maturity Criteria for Security Practices (SP)}}\\
        \cmidrule{2-6}
         & \multicolumn{2}{c}{
         {\centering\textbf{Quality of Implementation}}} &  \multicolumn{3}{c}{\textbf{Focus on DevOps Principles}}\\
         \cmidrule(r){2-3} \cmidrule(l){4-6}
         & \textit{SP execution} & \textit{SP structure: people, process, and technology aspects} & \textit{SP goals alignment with business first and customer focus principles} & \textit{SP continuous improvement with shift-left, automation, and continuous everything principles} & \textit{SP enables collaboration through continuous learning, experimentation, and feedback}\\
        \midrule
        \textbf{\textsf{0. Sit}} & \multicolumn{2}{l}{SP is not executed.}& & &    \\
        \midrule
        \textbf{\textsf{1. Crawl}} & \multicolumn{2}{l}{SP is executed in an ad-hoc manner.} & \multicolumn{3}{l}{SP's goals remain undefined.} \\
        \midrule
        \textbf{\textsf{2. Walk}} &  SP is executed: a) following defined elements (inputs/outputs),  b) at least for business-relevant projects & SP elements are somehow defined. For example, people: roles; or process: workflow; or technology: tools to support workflow.& SP's goal is to fulfill a requirement typically demanded by external stakeholders, for example, quality teams, or auditors. & SP is implemented: - mostly to the right - somehow automated - somehow iteratively / incrementally. & SP outcome is somehow used as input for other engineering practices. This is primarily driven by individuals, or isolated teams \\
        \midrule
        \textbf{\textsf{3. Run}}&  {SP is executed: a) following a defined structure (inputs/outputs/flow), and b) consistently, for most of the products, projects, and components.} & SP structure is well-defined. SP is defined as a standard practice. & SP's goal is to fulfil business and customer needs. The mindset is to implement explicit requirements. & SP improvement is visible through (i) a shift-left approach, (ii) automation, or (iii) iterative/incremental outcomes. 
        & SP outcome is often used as input for other engineering practices. This is a consequence of teams interacting for example retrospectives between Dev-Sec and Ops-Sec.\\
        \midrule   
         \begin{itemize}[label={},noitemsep,leftmargin=0pt,topsep=0pt,partopsep=0pt]
         \item\textbf{\textsf{4. Jump}}
         \end{itemize}
         & \begin{itemize}[label={},noitemsep,leftmargin=0pt,topsep=0pt,partopsep=0pt]
        \item SP is executed: a) following the defined structure, b) consistently for all products, projects, and components, and c) consistently by all teams \end{itemize}
        & \begin{itemize}[label={},noitemsep,leftmargin=0pt,topsep=0pt,partopsep=0pt]
          \item SP's structure is well-known to the team. SP is perceived as a good/best practice. Teams understand the benefits and identify the drawbacks of the practice's structure. \end{itemize} 
        & \begin{itemize}[label={},noitemsep,leftmargin=0pt,topsep=0pt,partopsep=0pt]
          \item SP's goal is to reduce risk. The mindset is "risk reduction provides value to the customer or business". \end{itemize} 
        & \begin{itemize}[label={},noitemsep,leftmargin=0pt,topsep=0pt,partopsep=0pt]
          \item SP's improvement is tracked and made visible to teams, who can provide feedback on the shift-left approach and aautomation roadmap \end{itemize}  
        & \begin{itemize}[label={},noitemsep,leftmargin=0pt,topsep=0pt,partopsep=0pt] 
        \item SP outcomes are used as input for other engineering practices. 
        \item SP prerequisites are linked to the outcome of other engineering practices. Dev-Sec-Ops teams track SP's outcomes vs. risk reduction. 
        \item Cross-functional teams learn from tracking dashboards. \end{itemize}\\
        \midrule
        \begin{itemize}[label={},noitemsep,leftmargin=0pt,topsep=0pt,partopsep=0pt]
        \item \textbf{\textsf{5. Fly}} \end{itemize}
        & \begin{itemize}[label={},noitemsep,leftmargin=0pt,topsep=0pt,partopsep=0pt]
        \item SP is engrained in the teams and the behavior of business stakeholders. \end{itemize}
        & \begin{itemize}[label={},noitemsep,leftmargin=0pt,topsep=0pt,partopsep=0pt]
        \item SP structure is well-known to teams and related business and management. 
        \item SP is considered as a best practice. 
        \item Teams and business stakeholders are aware of the benefits/drawbacks of the practice's structure.\end{itemize}   
        & \begin{itemize}[label={},noitemsep,leftmargin=0pt,topsep=0pt,partopsep=0pt]
        \item SP's goal is to reduce risk effectively and efficiently. The mindset is "effective and efficient risk reduction produces value for the customer and business". \end{itemize} & 
        \begin{itemize}[label={},noitemsep,leftmargin=0pt,topsep=0pt,partopsep=0pt]
        \item SP is constantly optimized as a result of a culture of continuous improvement.
        \item SP is implemented as early as possible in the product lifecycle
        \item SP is automated as fully as possible
        \item SP follows the continuous everything approach
        \end{itemize} 
        & \begin{itemize}[label={},noitemsep,leftmargin=0pt,topsep=0pt,partopsep=0pt]
        \item SP prerequisites and outcomes are well connected to engineering practices. \item SP influence on the software engineering process is tracked and allows one to predict and adapt the SP towards efficient risk reduction. 
        \item Cross-functional teams perform well with continuous experimentation practices involving individuals from DevSecOps as needed. \end{itemize} \\
        \midrule
\end{tabularx}
\caption{Maturity Scale: Criteria employed to rate security practices (SP) and generate \acrfull{refamr}.}
\label{tab:maturityScale} 
\end{table}

}

\end{document}